\documentclass[aps,prd,preprint,preprintnumbers,amsmath,amssymb,groupedaddress,nofootinbib,epic]{revtex4}
\pdfoutput=1
\usepackage{color}
\definecolor{darkblue}{rgb}{0.1,0.1,.1}
\usepackage[colorlinks, linkcolor=darkblue, citecolor=darkblue, urlcolor=darkblue, linktocpage]{hyperref}
\usepackage{natbib}
\setcitestyle{square, comma, compress,numbers}
\usepackage[]{amsmath}
\usepackage[]{graphicx}
\usepackage[]{epic}
\usepackage[]{eepic}
\usepackage{amscd}
\usepackage[all,cmtip]{xy}
\usepackage{mathrsfs}

\usepackage{simplewick}
\usepackage{changepage}
\usepackage{bm}
\usepackage [latin1]{inputenc}

\usepackage{amsfonts}
\usepackage{placeins}

\newcommand{\overbar}[1]{\mkern 1.5mu\overline{\mkern-0mu#1\mkern-3mu}\mkern 1.5mu}

\newcounter{RomanNumber}

\newcommand{\bea}{\begin{eqnarray}}
\newcommand{\eea}{\end{eqnarray}}
\newcommand{\beq}{\begin{equation}}
\newcommand{\eeq}{\end{equation}}

\def\<{\langle}
\def\>{\rangle}

\def\nn{\nonumber}

\def\cO {{\cal O}}

\begin{document}


\title{Searching for gauge theories with the conformal bootstrap}

\author{ Zhijin Li}
\email{lizhijin18@gmail.com}
\author{David Poland}
\email{david.poland@yale.edu}
\vspace{0.1cm}
\affiliation{
${}$
Department of Physics, Yale University, New Haven, CT 06511 \vspace{2cm}}

\begin{abstract}
\vspace{3mm}
Infrared fixed points of gauge theories provide intriguing targets for the modern conformal bootstrap program. In this work we provide some preliminary evidence that a family of gauged fermionic CFTs saturate bootstrap bounds and can potentially be solved with the conformal bootstrap. We start by considering the bootstrap for $SO(N)$ vector 4-point functions in general dimension $D$. In the large $N$ limit, upper bounds on the scaling dimensions of the lowest $SO(N)$ singlet and traceless symmetric scalars interpolate between two solutions at $\Delta =D/2-1$ and $\Delta =D-1$ via generalized free field theory. In 3D the critical $O(N)$ vector models are known to saturate the bootstrap bounds and correspond to the kinks approaching $\Delta =1/2$ at large $N$. We show that the bootstrap bounds also admit another infinite family of kinks ${\cal T}_D$, which at large $N$ approach solutions containing free fermion bilinears at $\Delta=D-1$ from below. The kinks ${\cal T}_D$ appear in general dimensions with a $D$-dependent critical $N^*$ below which the kink disappears. We also study relations between the bounds obtained from the bootstrap with $SO(N)$ vectors, $SU(N)$ fundamentals, and $SU(N)\times SU(N)$ bi-fundamentals. We provide a proof for the coincidence between bootstrap bounds with different global symmetries. We show evidence that the proper symmetries of the underlying theories of ${\cal T}_D$ are subgroups of $SO(N)$, and we speculate that the kinks ${\cal T}_D$ relate to the fixed points of gauge theories coupled to fermions. 

\end{abstract}

\maketitle

\newpage

\tableofcontents

\newpage
\section{Introduction}
\label{sec:introduction}
The modern conformal bootstrap \cite{Rattazzi:2008pe} provides a powerful nonperturbative approach to study higher dimensional conformal field theories (CFT). This method exploits general consistency conditions satisfied by all conformal theories to generate remarkably precise CFT data with rigorous control on the errors. This method is particularly useful for studying strongly-coupled conformal theories for which perturbative approaches are not applicable. Following some remarkable successes in the 3D critical Ising and $O(N)$ vector models \cite{ElShowk:2012ht, Kos:2013tga, El-Showk:2014dwa, Kos:2014bka, Kos:2015mba, Kos:2016ysd} (and more recently \cite{Chester:2019ifh}), the conformal bootstrap has been used to tackle various types of CFTs in higher dimensions $D>2$ (see \cite{Poland:2018epd} for a review). Nevertheless, most CFTs that saturate bootstrap bounds obtained so far (particularly non-supersymmetric ones) are limited to theories without gauge interactions.\footnote{With supersymmetry the conformal bootstrap can benefit from supersymmetry-based analytical techniques, such as localization and chiral algebras, making it easier to constrain or even numerically solve supersymmetric CFTs with gauge interactions, see e.g.~\cite{Beem:2013qxa, Chester:2014fya, Beem:2014zpa, Chang:2017xmr, Chang:2017cdx, Cornagliotto:2017snu, Agmon:2017xes, Agmon:2019imm, Chang:2019dzt}.}

On the other hand, a large class of CFTs in higher dimensions are realized through gauge interactions. The physically interesting theories are usually strongly coupled and require non-perturbative approaches, such as lattice simulations, to study their infrared (IR) dynamics. Two classic examples are given by 3D Quantum Electrodynamics (QED$_3$) and 4D Quantum Chromodynamics (QCD$_4$), which have broad  applications in condensed matter systems and high energy physics. Low energy limits of the two theories include both conformal and chiral symmetry breaking phases depending on the flavor number. Near the critical flavor number the theories become strongly coupled and it turns out to be extremely challenging to determine their IR dynamics.

As a surprisingly powerful nonperturbative approach, the conformal bootstrap is expected to shed light on these profound strong coupling problems. In particular, the conformal bootstrap has been used to provide non-trivial constraints on the IR dynamics of QED$_3$ \cite{Nakayama:2016jhq, Chester:2016wrc, Li:2018lyb} and on those of 4D gauge theories \cite{Poland:2011ey, Caracciolo:2014cxa, Iha:2016ppj, Nakayama:2016knq, Karateev:2019pvw}. These constraints are helpful for answering certain questions relevant to the dynamics of gauge interactions. Nevertheless, they are not as strong as the results of the 3D critical Ising model, which appear to saturate the bootstrap bounds at a kink-like discontinuity and provide extremal solutions to the bootstrap equations. More generally, a kink-like discontinuity suggests the existence of a non-trivial solution to the crossing equation which may potentially be promoted to a full-fledged theory. This can be further tested by exploiting the consistency conditions with mixed correlators under suitable assumptions on the theory, which may allow one to isolate the solution. Therefore, we can heuristically consider a kink-like discontinuity to be a precursor to identifying a theory that can be solved with the conformal bootstrap.

In particular, some promising evidence towards bootstrapping 3D gauged CFTs without supersymmetry was found recently in \cite{Li:2018lyb}, which discovered a new family of kink-like discontinuities in the bootstrap bounds, with a possible relation to the infrared (IR) fixed points of QED$_3$. In the present work, we will extend this analysis and identify a new infinite family of kinks in the bootstrap bounds in general dimensions, which we conjecture to be related to full-fledged non-supersymmetric CFTs with gauge interactions. We will particularly focus on the interpretation of these kinks as they appear in the 4D bootstrap applied to 4-point functions of fermion bilinears.

In search of an infinite family of CFTs, such as fixed points of QED$_3$ or QCD$_4$, actually it is more illuminating to start with their large $N$ limit, since in this limit the theory is significantly simplified (QCD$_4$) or even solvable (QED$_3$). This is counter to the history of the numerical conformal bootstrap, in which the first numerical solution was obtained for the critical Ising model \cite{Rychkov:2009ij, ElShowk:2012ht, El-Showk:2014dwa} and then the critical $O(N)$ vector models \cite{Kos:2013tga, Kos:2015mba}. In \cite{Li:2018lyb} an infinite family of kinks (${\cal T}_{3D}$) beyond the well-known critical $O(N)$ vector model ones were discovered in 3D bootstrap bounds.
Combining the results in \cite{Li:2018lyb} with the earlier bootstrap kinks connected to the 3D critical $O(N)$ vector model \cite{Kos:2013tga}, it gives a rather interesting pattern of kinks in the 3D $SO(N)$ vector bootstrap:

{\it There are two infinite families of kinks in the 3D $SO(N)$ vector bootstrap, which respectively approach solutions to the crossing equation with a (scalar) $SO(N)$ vector at $\Delta=1/2$ and $\Delta=2$, both containing a series of conserved higher spin currents. The kinks approaching $\Delta=2$ have an additional fine structure consisting of two nearby kinks at each value of $N$ above a critical value $N^*\simeq6$.}

\begin{figure}[!htb]
\includegraphics[scale=0.8]{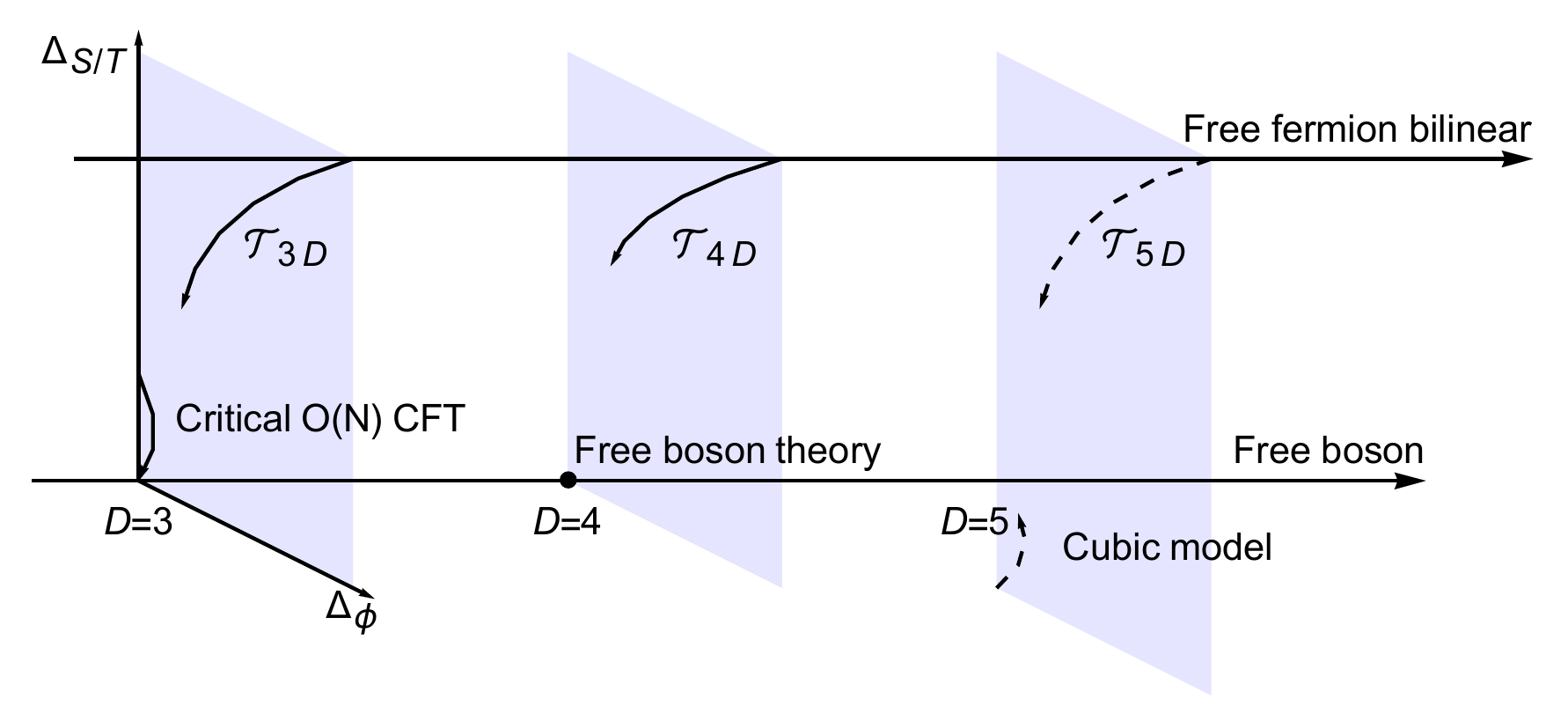}
 \begin{flushright}
\caption{ A sketch of the kink-like discontinuities in the $SO(N)$ vector bootstrap bounds on the scaling dimensions of $SO(N)$ singlet/traceless symmetric scalars in dimensions $D=3,4,5$. The curves ending in arrows denote the positions of the kinks in the bootstrap bounds as one moves from $N=\infty$ to the critical value $N^*$. There are two families of kinks which correspond to deformations of free boson and free fermion theories respectively. In 4D the critical $O(N)$ vector models become free and there is no analogous kink in the bootstrap bound. In 5D the family of kinks approaching the free boson theory correspond to 5D $O(N)$ models, which are perturbatively stable and unitary above the critical value $N^*$. However, unitarity in these models is violated by non-perturbative effects. The kinks approaching free fermion bilinears ($\cal T_{D}$) appear in general dimensions with a D-dependent critical value $N^*$, which we speculate are related to fermionic gauged CFTs. They are the main objects of this study. } \label{total}
\end{flushright}
\end{figure}

The large $N$ behavior of this new set of kinks is quite enlightening when considering the interpretation in terms of an underlying Lagrangian description. In higher dimensions $(D>2)$, a theory with conserved higher spin currents is essentially free \cite{Maldacena:2011jn, Boulanger:2013zza, Alba:2015upa}. In the large $N$ limit the new family of kinks approach free fermion theory from below.\footnote{One may wonder how a scalar $SO(N)$ vector appears in a free fermion theory. Actually there is a symmetry enhancement  in the bootstrap results due to the bootstrap algorithm. We will discuss this phenomenon in section \ref{sec:cbz}.} In 3D the structure at large $N$ seems to cleanly resolve into {\it two} closely separated kinks and the the scaling dimensions of non-singlet fermion bilinears at finite $N$ nicely agree with the $1/N$ corrections arising in QED$_3$ and QED$_3$-GNY (Gross-Neveu-Yukawa) models. This leads to a conjecture that the kinks at all $N$ relate to the IR fixed points of QED$_3$ and QED$_3^* = $ QED$_3$-GNY, and they merge at the critical flavor number $N^*$!

It is natural to ask if we can find similar patterns in the bootstrap bounds beyond 3D. In 4D, there are no interacting IR fixed points in the $O(N)$ vector models and the corresponding kinks disappear in the 4D bootstrap results.\footnote{Actually it has been suggested in \cite{Bond:2018oco} that no weakly coupled fixed point can be generated in 4D without gauge interactions.} On the other hand, the IR fixed points can be realized in asymptotically free Yang-Mills theories coupled to massless fermions, known as Caswell-Banks-Zaks fixed points  (CBZ) \cite{Caswell:1974gg, Banks:1981nn}.\footnote{In this paper, we use ``CBZ fixed points" to denote the CFTs in whole conformal window. We note that in certain terminology ``CBZ fixed points" refers to CFTs near the upper bound of the conformal window only. }
To realize the CBZ fixed points, the number of massless fermions must be inside of an interval, namely the  ``conformal window". The upper limit of the conformal window is reached when asymptotic freedom is lost, while below the lower bound chiral symmetry breaking and confinement will be triggered in the low energy limit. The CBZ fixed points play important roles in possible scenarios of physics beyond the standard model, and provide classic examples of CFTs with strongly-coupled gauge interactions. They have also been extensively studied using lattice simulations. General bounds on the CFT data of CBZ fixed points can be obtained through the conformal bootstrap, though the bounds obtained so far are fairly weak \cite{Poland:2011ey, Caracciolo:2014cxa, Iha:2016ppj, Nakayama:2016knq, Karateev:2019pvw}.

An extremely interesting question is whether the CBZ fixed points can saturate bootstrap bounds at kink-like discontinuities, an indication that the theories could potentially be isolated and numerically solved using the conformal bootstrap. Surprisingly, we do find a family of kinks (${\cal T}_{4D}$) in the 4D bootstrap bounds, as briefly sketched out in Figure \ref{total}, though we do not know their putative Lagrangian descriptions yet. In this work we will study the kink-like discontinuities ${\cal T}_{4D}$ based on the scenario mentioned before and discuss their possible relations with the CBZ fixed points.

This work is organized as follows. In section~\ref{sec:3d} we review results on the new kinks ${\cal T}_{3D}$ from the 3D $SO(N)$ vector bootstrap, their relation to the $SU(N)$ adjoint bootstrap, and their possible connections to the IR fixed points of QED$_3$. In section~\ref{sec:4d} we move to the 4D $SO(N)$ vector bootstrap and study the behavior of the kinks both in the large $N$ limit and near the apparent critical value $N^*$.  In section~\ref{sec:cbz} we study the relation between the 4D $SO(N)$ vector and $SU(N_f)\times SU(N_f)$ bi-fundamental bootstrap and give a proof of the coincidence between the the bootstrap bounds with different global symmetries. We further discuss the possible relation between the bootstrap results and the CBZ fixed points. In section~\ref{sec:5d} we describe a similar bootstrap study in 5D. We conclude in section~\ref{sec:discussion} and discuss future work towards bootstrapping the fixed points of gauge theories.

\FloatBarrier\section{Kinks in the 3D $SO(N)$ vector bootstrap}
\label{sec:3d}
Conformal QED$_3$ \cite{Appelquist:1988sr} provides arguably the simplest examples of CFTs realized as fixed points of gauge theories in higher dimensions $D\geqslant3$.
In its standard version, QED$_3$  is a $U(1)$ gauge theory coupled to $N_f$ flavors of two-component Dirac fermions $\psi_i$. The IR phase of this model is surprisingly fertile: its low-energy limit can realize a conformal phase as an IR fixed point of the RG flow, chiral symmetry breaking, or even confinement depending on the flavor number $N_f$ \cite{Polyakov:1975rs, Polyakov:1976fu, Appelquist:1988sr, Pisarski:1984dj, Appelquist:1986fd}. Therefore QED$_3$ provides an appropriate playground to study these profound phenomena at strong coupling. Likewise, conformal QED$_3$ provides an ideal target for the conformal bootstrap to learn how to study CFTs with gauge interactions. In bootstrap studies, one primarily focuses on gauge-invariant operators. The leading gauge-invariant operators in QED$_3$ are the fermion bilinears $\bar{\psi}_i\psi^j$ and the monopole operators, both of which furnish non-trivial representations of the flavor symmetry $SU(N_f)$.
In \cite{Chester:2016wrc}, the authors applied the conformal bootstrap to conformal QED$_3$ with an emphasis on the monopole operators, which are characteristic of QED$_3$. The fermion bilinear 4-point correlator was more recently bootstrapped in \cite{Li:2018lyb}. The results show interesting relations with conformal QED$_3$ in several aspects.

\begin{figure}[!htb]
\includegraphics[scale=0.7]{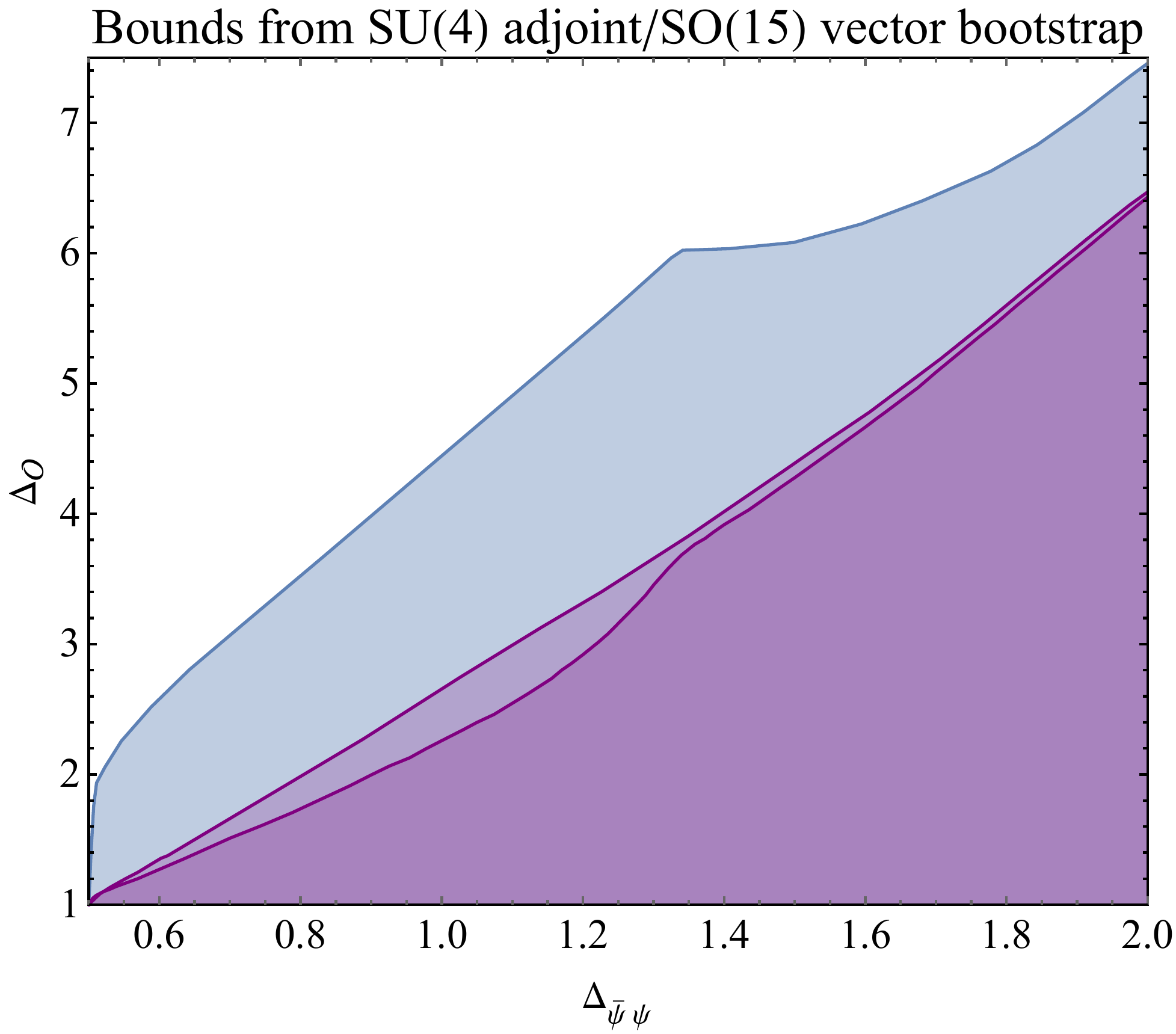}
 \begin{flushright}
\caption{Bounds on the scaling dimensions of the lowest scalars in the $SU(4)$ singlet (blue line) and  $(T,\bar{T})$ representation (higher purple line) appearing in the OPE $\cO_{\text{adj}}\times \cO_{\text{adj}}$, and a bound on scaling dimension of the lowest scalar in the $SO(15)$ traceless symmetric representation (lower purple line) appearing in the OPE $\phi_i \times \phi_j$. The $SU(4)$ singlet bound coincides with the singlet bound obtained from the $SO(15)$ vector bootstrap. The kink in the singlet bound near $(0.5, 2)$ relates to the critical $O(15)$ vector model. In addition, there is a new prominent kink. While not easy to resolve on this plot, near the second kink there is an interesting fine structure as shown in Figure 3 of \cite{Li:2018lyb}. } \label{3DO15}
\end{flushright}
\end{figure}

In QED$_3$, the fermion bilinears $\cO_{\text{adj}} \sim\bar{\psi}_i\psi^j$ transform in the adjoint representation of the flavor symmetry $SU(N_f)$. Applying semidefinite programming methods using SDPB \cite{Simmons-Duffin:2015qma, Landry:2019qug}, the 4-point correlator of fermion bilinears can be used to generate rigorous bounds on CFT data. 
Surprisingly, the bound on scaling dimension of the leading singlet scalar $\cO_S$ appearing in the OPE $\cO_{\text{adj}}\times \cO_{\text{adj}} \sim 1+ \cO_S+\cdots$ coincides with the bound obtained from the $SO(N)$ vector bootstrap, given $N=N_f^2-1$! Similar coincidences among the bootstrap bounds with different global symmetries have been observed before \cite{Poland:2011ey, Nakayama:2017vdd}. 
There are several non-singlet scalars $\cO_{R}$ appearing in the OPE $ \cO_{R}\in \cO_{\text{adj}}\times \cO_{\text{adj}} $ and their bounds depend on their representations. Without additional assumptions their upper bounds are higher (weaker) than that of the $SO(N)$ traceless symmetric scalar. On the other hand, they all become identical if these non-singlet scalars are restricted to have the same scaling dimension. In a physical theory, this would only hold in the large $N$ limit when the composite operators appearing in the OPE are factorized. With finite $N$ the assumption is true at leading order and is violated by $1/N$ corrections. Due to these coincidences of the bounds, it is subtle to determine the true global symmetry of a putative theory saturating the bounds. This problem will be studied further in section \ref{sec:cbz}.

Here we primarily wish to highlight the new family of prominent kinks appearing in the bootstrap bounds, see Figure \ref{3DO15} for an example with $N_f=4$. Bounds on the scaling dimensions of the lowest scalars in the $SU(4)$ singlet and  $(T,\bar{T})$\footnote{Operators in this representation carry two fundamental and two anti-fundamental indices, both of which are symmetrized.}  representations are shown in the figure.
The kinks remain in the bounds at larger $N_f$, and they approach  $\Delta_{\text{adj}}=2$ from below in the limit $N_f\rightarrow\infty$. Meanwhile, the bound on the singlet scaling dimension becomes weaker and finally disappears when $N_f\rightarrow\infty$, while the scaling dimension of the $SO(N)$ traceless symmetric scalar has a scaling dimension $\Delta=4$ near the kink. 

Using the extremal functional method \cite{Poland:2011ey, ElShowk:2012hu, El-Showk:2014dwa}, we can obtain a picture of the spectrum near the kink.  In the large $N_f$ limit, there appears both a series of conserved higher-spin currents as well as double-trace operators from generalized free field theory (see section \ref{largeNEFM} for a similar analysis in 4D), suggesting that the large $N_f$ spectrum corresponds to a mixture between generalized free field theory and a free theory associated with a non-singlet scalar of scaling dimension $2$, i.e., a free fermion theory. 

Consequently, the kinks at finite $N_f$, if they correspond to full-fledged theories, are expected to relate to interacting perturbations of free fermion theory!\footnote{Like the result with $N_f\rightarrow \infty$, at large but finite $N_f$, it is possible that the extremal solution at the kink still picks out a mixture between the underlying theory and a generalized free field theory. Therefore, without imposing a finite central charge, the kinks may relate to but perhaps cannot be directly identified with a (local) physical theory.}
 A well-known example of such a deformation of free fermion theory is the Gross-Neveu model \cite{Gross:1974jv}, which is typically realized as a UV fixed point containing a four-fermion interaction, or equivalently as an IR fixed point containing a Yukawa coupling (the Gross-Neveu-Yukawa model). However, in this non-gauged interacting theory the non-singlet fermion bilinears have positive anomalous dimension (see e.g.~\cite{Iliesiu:2015qra}). In the large $N_f$ limit, their scaling dimension approaches $\Delta_{\text{adj}}=2$ from above.
It turns out that the large $N_f$ behavior of  $\Delta_{\text{adj}}$ shown in the numerical results is instead consistent with the large $N_f$ perturbative expansions of QED$_3$ and QED$_3$-GNY \cite{Xu08, Chester:2016ref, Kotikov:2016prf, Gracey:1993sn, Gracey:2018fwq, Boyack:2018zfx}, indicating that the underlying theories of the new kinks may be related to conformal QED$_3$.

A particular advantage of the conformal bootstrap is that it works nicely no matter how strongly coupled the theory is. In QED$_3$, it is believed that there is a critical flavor number $N_f^*$, below which the theory runs into a chiral symmetry breaking phase in the low-energy limit. Near the critical flavor number the theory is strongly coupled and the value of $N^*_f$ is still under debate.
According to the proposed connection between the new family of kinks and conformal QED$_3$, its behavior at small $N_f$ could help us to estimate $N_f^*$.
The results in \cite{Li:2018lyb}  show that the kinks persist for $N_f\geqslant3$,\footnote{QED$_3$ with an odd flavor number of two-component Dirac fermions has a parity anomaly. Here we interpret odd $N_f$ as an analytical continuation of the CFT data while ignoring the parity anomaly. It will be interesting in the future to explore the implications of imposing parity symmetry in the mixed correlator bootstrap.} giving evidence that $N_f^*=2$.

Moreover, the results support the merger and annihilation mechanism \cite{10.1143/PTP.105.809,PhysRevB.71.184519,Gies:2005as,Kaplan:2009kr, Gorbenko:2018ncu}, through which the IR fixed point of QED$_3$ merges with the QED$_3$-GNY model and disappears near $N_f^*$.  The merger and annihilation mechanism is suggested to be triggered when an $SU(N_f)$ singlet four-fermion operator crosses marginality $\Delta_{(\bar{\psi}\psi)^2}=3$.\footnote{Note that the IR fixed point does not necessarily merge with another UV fixed point and disappear when a four-fermion operator crosses marginality. It is possible that two lines of fixed points cross instead of merge. In conformal QED$_3$, we observe a relevant four-fermion operator in a non-singlet representation of the flavor symmetry $SU(N_f)$, while the bound still shows a prominent kink. In the future we hope to provide a more detailed study on loss of conformality of fermionic gauge theories using the conformal bootstrap, both in 3D and higher dimensions. We thank S. Rychkov for insightful discussions on the mechanisms by which conformality can be lost.} Below $N_f^*$, the relevant four-fermion interaction is expected to generate an RG flow to the phase with chiral symmetry breaking, while the physics in this region goes beyond the reach of the conformal bootstrap. For the QED$_3$-GNY model with flavor number $N_f=2$, there is evidence supporting an $SO(5)$ symmetry enhancement in the IR phase \cite{Nahum:2015jya, Nahum:2015vka}, while it is questionable if it relates to a unitary CFT based on previous bootstrap studies \cite{Nakayama:2016jhq, DSD, Ili2018, Poland:2018epd}. Bootstrap results in \cite{Li:2018lyb} suggest that the putative CFT with enhanced $SO(5)$ symmetry is likely to have an $N_f$ just below the conformal window in 3D.\footnote{A dimensional continuation of this theory in the context of a $D=2+\epsilon$ dimensional nonlinear sigma model was studied in \cite{Ma:2019ysf, Nahum:2019fjw}, suggesting that conformality is lost at $D\simeq 2.77$. In the numerical bootstrap one can also study the dimensional continuation of the $SO(5)$~\cite{Li:unpub} or $SO(4+\epsilon)$~\cite{He:2020azu} vector bounds. In these cases the sharp kink seems to disappear near $D \simeq 2.7-2.8$.}

The 3D bootstrap results show promising evidence that the bounds can access CFTs perturbed from free fermion theory through $U(1)$ gauge interactions. It is tempting to ask if we can get similar results in 4D and even higher dimensions. Although gauge dynamics in 4D are quite different from those of 3D, on the conformal bootstrap side the spacetime dimension D is just a parameter in the implementation, and it is straightforward to apply a similar analysis to CFTs with $D\geqslant4$.

\FloatBarrier\section{Kinks in the 4D $SO(N)$ vector bootstrap}
\label{sec:4d}
In this section we show some results from the 4D $SO(N)$ vector bootstrap with an emphasis on the new family of kinks that approach free fermion theory in the large $N$ limit. As discussed above, the $SO(N)$ vector bootstrap results (including the kink-like discontinuities) actually coincide with those with different global symmetries, for instance $SU(N_f)\times SU(N_f)$ symmetry given $N=2N_f^2$. In consequence, the putative full-fledged theories connected to the kinks, if they exist, do not necessarily have $SO(N)$ global symmetry. Instead, the proper global symmetry of the theory could be a subgroup of $SO(N)$. We'll discuss more about this symmetry enhancement phenomena in section \ref{sec:cbz}, but first we wish to show several interesting properties of the bootstrap results.

In the $SO(N)$ vector bootstrap, one focuses on the 4-point correlator of the $SO(N)$ vector $\phi_i$: $\langle \phi_i\phi_j\phi_k\phi_l\rangle$. Its crossing equation includes three channels
\bea
\sum_{S^+}\lambda_\cO^2 \left( \begin{array}{cc}  0  \\ F_{\Delta,\ell} \\ H_{\Delta,\ell} \\
\end{array} \right)+\sum_{T^+}\lambda_\cO^2 \left( \begin{array}{cc}  F_{\Delta,\ell}  \\ (1-\frac{2}{N}) F_{\Delta,\ell} \\ -(1+\frac{2}{N})H_{\Delta,\ell} \\
\end{array} \right)+\sum_{A^-}\lambda_\cO^2 \left( \begin{array}{cc}  -F_{\Delta,\ell}  \\  F_{\Delta,\ell} \\ -H_{\Delta,\ell} \\
\end{array} \right)=0, \label{ONce}
\eea
where $(S, T, A)$ denote singlet, traceless symmetric and anti-symmetric representations of $SO(N)$ symmetry. The superscript signs in $X^{\pm}$ denote the even/odd spins that can appear in the channel $X$. Here $F_{\Delta,\ell}/H_{\Delta,\ell}$ are the $(u,v)$ symmetrized/anti-symmetrized functions:
\bea
F_{\Delta,\ell}=v^{\Delta_\phi}g_{\Delta,\ell}(u,v)-u^{\Delta_\phi}g_{\Delta,\ell}(v,u), \label{Fblock}\\
H_{\Delta,\ell}=v^{\Delta_\phi}g_{\Delta,\ell}(u,v)+u^{\Delta_\phi}g_{\Delta,\ell}(v,u), \label{Hblock}
\eea
where $g_{\Delta,\ell}(u,v)$ is a conformal block \cite{Dolan:2000ut, Dolan:2003hv, Dolan:2011dv}. Numerical computations are carried out using the code \cite{sboot, DPcode} which calls the semi-definite programming solver SDPB \cite{Simmons-Duffin:2015qma, Landry:2019qug}.  

Before proceeding, let us note that in this correlator the bootstrap implementations for $O(N)$ and $SO(N)$ symmetries would be indistinguishable. Thus, while CFTs corresponding to solutions of these equations may have a full $O(N)$ symmetry, we cannot determine this without probing a larger system of correlators.

\FloatBarrier\subsection{Bounds on the scaling dimensions at small $N$}

\begin{figure}[!htb]
\includegraphics[scale=0.7]{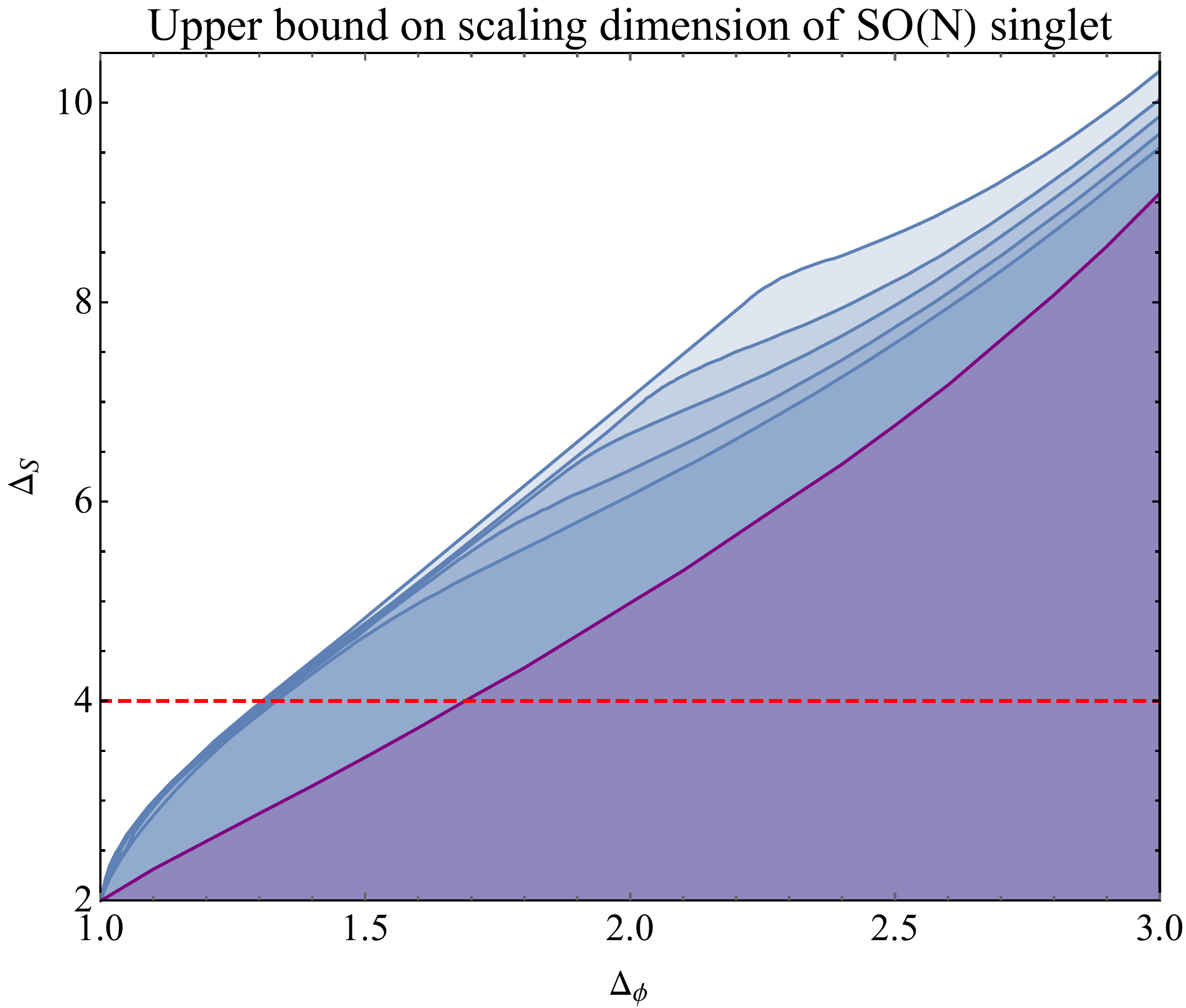}
 \begin{flushright}
\caption{ Upper bounds (blue lines) on the scaling dimensions of the lowest scalars in the singlet sector appearing in the $\phi_i\times\phi_j$ OPE, with $N=14,18,24,32,50$. The bound increases monotonically with $N$. The purple line gives the upper bound on the scaling dimension of the $SO(18)$ traceless symmetric scalar. Bounds on the $SO(N)$ traceless symmetric scalar are largely degenerate for small $N$ and the bound for $N=14$ is very close to the bound shown in the figure. The dashed red line is the marginality condition $\Delta=4$.
The bounds are computed with maximum derivative order $\Lambda=31$. } \label{4DONs}
\end{flushright}
\end{figure}

Bounds on the scaling dimension of the lowest scalar in the singlet sector for $N=14, 18, 24, 32, 50$ are presented in Figure \ref{4DONs}.  
For the 4D $SO(N)$ vector $\phi$, the bounds are smooth near its unitary bound $\Delta=1$. In contrast,  as shown in Figure \ref{3DO15}, the singlet bounds obtained from the 3D $SO(N)$ vector bootstrap have sharp kinks near the unitary bound $\Delta=0.5$, corresponding to the 3D critical $O(N)$ vector models. This is consistent with the fact that the IR fixed points of the $O(N)$ vector models merge with the free boson fixed points in 4D.

Interestingly, the bounds show notable jumps and kink-like discontinuities for $N=32, 50$ near $\Delta_{\phi}\sim 2, 2.2$. The kinks become sharper at larger $N$, see e.g.~the $SO(288)$ singlet bound in Figure \ref{SU(12)}. The kink becomes less sharp at $N=18$ and indistinguishable at $N=14$, potentially suggesting a critical number $N^*$   near these values. While suggestive, it is hard to determine the precise $N^*$ based on the smoothness of the bounds with the current numerical precision.

A sharper criterion for identifying the kink could help to better estimate the critical flavor number $N^*$. One way to do this is to look for transitions in the extremal spectrum near the kink location. We have checked the spectra in the extremal solutions near the kink, and see that for $N>N^*$ where there is a distinguishable kink, the kink is actually accompanied by an operator in the singlet sector decoupling from the spectrum, similar to the phenomenon observed in \cite{El-Showk:2014dwa}. Specifically, in the $N=18$ bootstrap bound with $\Lambda=31$,  the operator decoupling occurs near $\Delta_\phi\simeq 1.75$. Though the bootstrap bound becomes smooth for $N=14$, the operator decoupling still appears near $\Delta_\phi\simeq 1.45$. On the other hand, we do not see any similar operator decoupling in the $SO(N)$ bootstrap bounds for $N\leqslant13$. If the operator decoupling is a signal of an underlying CFT, it would suggest $N^*=14$. 
However, different from the critical 3D Ising model \cite{El-Showk:2014dwa}, the operators that we observe decoupling from spectrum near the family of kinks $\cal T_{D}$ usually have high scaling dimensions ($\Delta \sim 17$ for $N=18$, $\Lambda=31$) which are also not very well converged. So the precise connection between the operator decoupling phenomena and the kinks could be modified at higher numerical precision and should be taken with a grain of salt.

\begin{figure}[!htb]
\includegraphics[scale=0.5]{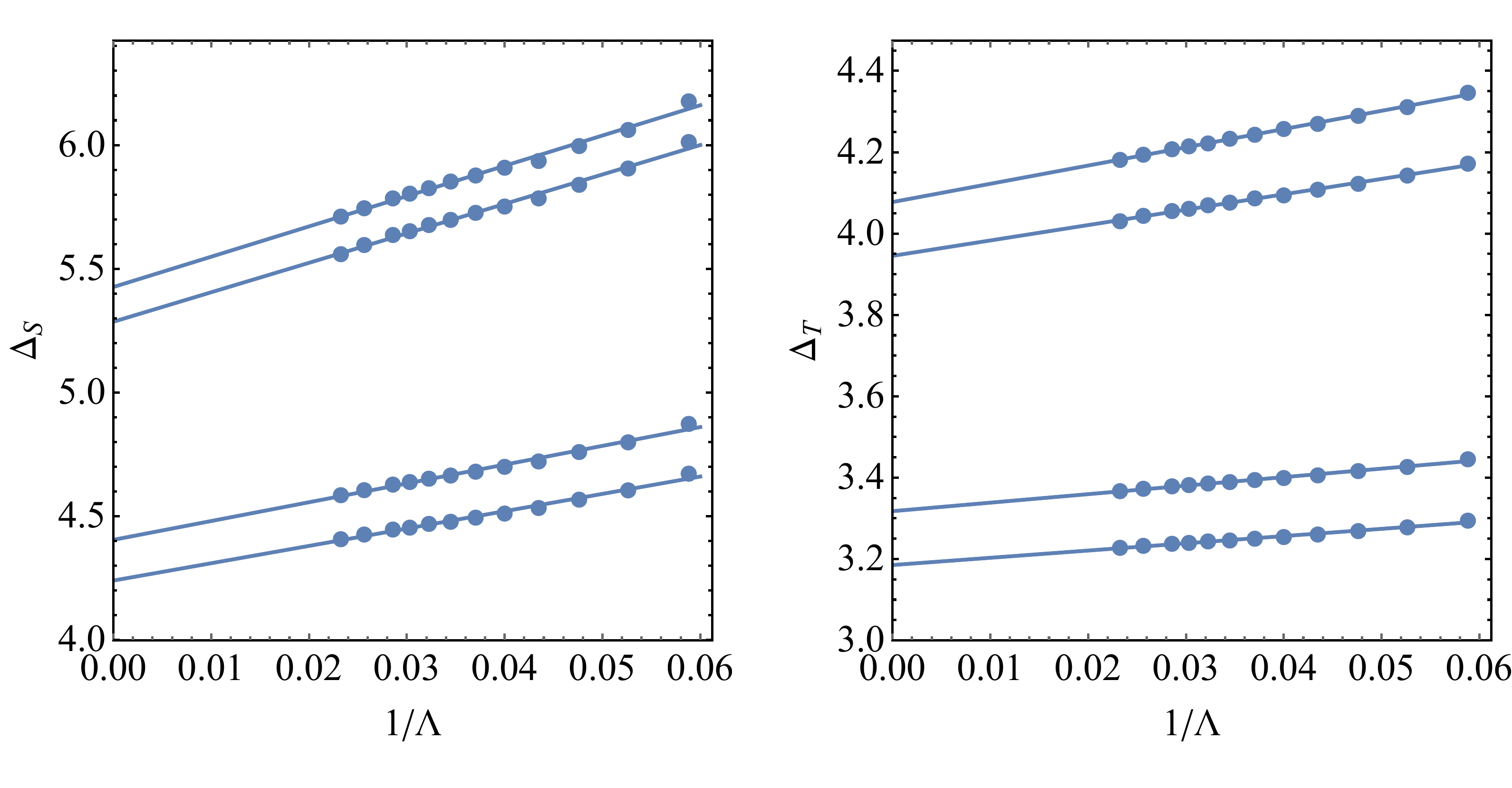}
 \begin{flushright}
\caption{Left panel: from top to bottom, linear extrapolations of the 4D $SO(N)$ singlet scalar upper bounds with $(N=18, \, \Delta_\phi=1.80), ~(N=18, \,\Delta_\phi=1.75),~ (N=14, \,\Delta_\phi=1.50), ~(N=14, \,\Delta_\phi=1.45)$. Right panel: linear extrapolations of the $SO(N)$ traceless symmetric scalar upper bounds with the same order as in the left panel.} \label{SOextrp}
\end{flushright}
\end{figure}

At small $N$ the bounds on the scaling dimensions of the lowest scalars in the traceless symmetric sector ($T$) seem to be featureless, and they do not change much for different $N\sim20$.
However, there is interesting information hidden in the smooth bound. Let us compare the bounds on the scaling dimensions of the singlet and traceless symmetric scalars. The series of kinks seem to disappear at a certain $N^*$ below $N=18$, and for $N=18$, there is a mild kink and the scaling dimension of the $SO(N)$ vector is roughly estimated in the range $\Delta_\phi\in (1.7, 1.8)$.
In the bound on the traceless symmetric scalar, near $\Delta_\phi\sim 1.7$ the upper bound approaches $\Delta_T=4$, i.e., the lowest traceless symmetric scalar cannot be irrelevant.   The relation between the upper bounds on the $SO(N)$ singlet and traceless symmetric scalars can be further studied using the extremal functional method. In the extremal solution at the $SO(N)$ singlet upper bound, we see that the lowest traceless symmetric scalar coincides with the upper bound on the lowest traceless symmetric scalar, so in the extremal spectrum it is becoming marginal.

The bootstrap bounds in Figure \ref{4DONs} are still not well converged even at $\Lambda\sim31$. In Figure \ref{SOextrp} we estimate the large $\Lambda$ behavior of the upper bounds using a linear extrapolation.  The results suggest
a nice linear relation $\Delta_{S/T} \propto 1/\Lambda$. Remarkably, for $N=18$, below which the bootstrap bound becomes relatively smooth,
the traceless symmetric scalar at the kink location $\Delta_{\phi} \sim (1.75,1.8)$ approaches marginality, $\Delta_T \simeq 4$,\footnote{Here we implicitly assume the $N=18$ kink stays in the region near $\Delta_\phi\sim(1.75, 1.8)$ at larger $\Lambda$, and similarly $\Delta_\phi \sim(1.45, 1.50)$ for $N=14$ (where we see a singlet scalar decouple in the spectrum). This is based on the observation that the $\Delta_\phi$ of the kink, compared with $\Delta_{S,T}$, is more stable to increasing $\Lambda$.} while the lowest singlet scalar stays irrelevant. In contrast, for $N=14$, near $\Delta_\phi \sim (1.45, 1.50)$ the lowest traceless symmetric scalar is relevant while the lowest singlet scalar is close to being marginal. To summarize, depending on the value of critical flavor number $N^*$, the disappearance of the kink is either accompanied by a marginal singlet scalar or traceless symmetric scalar.

In 3D, a similar family of kinks seems to disappear when the singlet scaling dimension approaches $\Delta_S=3$ and becomes marginal \cite{Li:2018lyb}. It will be interesting to better determine if the 4D kinks disappear as the singlet bound crosses marginality, or if there is a more exotic scenario of the symmetric tensor approaching marginality at $N^*$.  However, we leave a better determination of $N^*$ and which operator approaches marginality to a future study.  If the kinks do relate to full-fledged CFTs, this could provide strong evidence on the mechanism by which conformality is lost. We will give additional discussion on this point after clarifying several aspects of the 4D results.

\FloatBarrier\subsection{Bounds on the scaling dimensions at large $N$}  \label{largeNEFM}
The kinks shown in Figure \ref{4DONs} persist with larger $N$,  and approach the position $\Delta_\phi=3$ in the large $N$ limit. The upper bound on the scaling dimension of the singlet scalar becomes weaker at larger $N$ and disappears as $N\rightarrow\infty$. In contrast, the upper bound on the scaling dimension of the traceless symmetric scalar gets stronger at larger $N$. In the large $N$ limit the bound in the region $\Delta_{\phi}< 3$ is saturated by generalized free field theory. This was previously conjectured in \cite{Nakayama:2016knq}.
On the other hand, at precisely $\Delta_{\phi}=3$ there is another solution to the crossing equation given by free fermion theory, coinciding with a sharp transition in the bootstrap bound at $\Delta_{\phi}=3$. 
The sharp transition at infinite $N$ in 3D was suggested to be related to the free fermion theory in \cite{Li:2018lyb}. In this section we will provide more
evidence on this relation in 4D. An analytical understanding of the underlying four-point correlator with $N=\infty$ and $\Delta_{\phi}=3$
can be found in the parallel work \cite{He:2020azu}.\footnote{The analytical solution was first described in the talk \cite{ningsutalk}.}

\begin{figure}[!htb]
\includegraphics[scale=0.7]{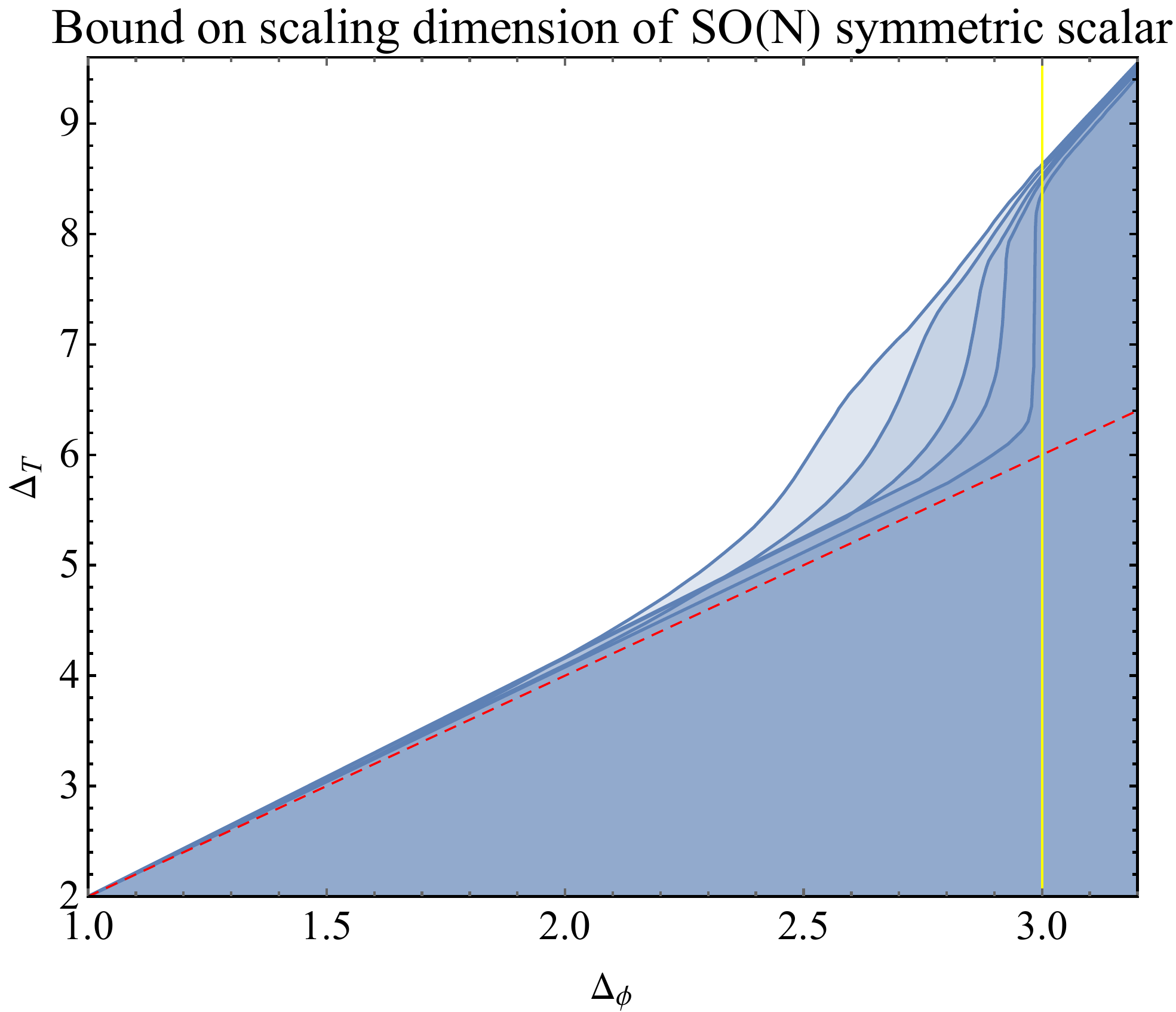}
 \begin{flushright}
\caption{Bounds on the scaling dimensions of the lowest $SO(N)$ traceless symmetric  scalars appearing in the $\phi_i\times \phi_j$ OPE, with $N=128, 288, 800, 1800, 20000$. The bound decreases monotonically with larger $N$. The red dashed line shows the relation $\Delta_T=2\Delta_\phi$ satisfied by generalized free field theory. The bound approaches generalized free field theory in the large $N$  limit and a kink/jump appears near the same $\Delta_\phi$ where the kink appears in the singlet bound. The bounds shown in the figure can be related to the bounds of the $SU(N_f)\times SU(N_f)$ ($N_f=8, 12, 20, 30, 100$) bi-fundamental bootstrap with suitable assumptions. } \label{4DONT}
\end{flushright}  
\end{figure}

Bounds on the scaling dimensions of the lowest $SO(N)$ traceless symmetric scalars appearing in the $\phi_i\times\phi_j$ OPE are shown in Figure \ref{4DONT} for  $N=128, 288, 800, 1800, 20000$.
At large $N$ and in the region with small $\Delta_\phi$, the bound is close to generalized free field theory, with a fixed relation between scaling dimensions of the $SO(N)$ symmetric scalar $\Delta_T$ and $SO(N)$ vector:  $\Delta_T=2\Delta_\phi$. The bounds show kinks/jumps which become sharper at large $N$ and less prominent at small $N$. The $x$-positions of the kinks/jumps, i.e., the scaling dimensions of the $SO(N)$ vectors $\Delta_\phi$, are close to the $x$-positions of the kinks in the singlet bounds (see e.g.~Figure \ref{SU(12)}).  In Figure \ref{4DONT}, we can observe an interesting property of these kink/jump locations:  the anomalous dimension of the $SO(N)$ vector at large $N$ seems to scale as
\bea
\gamma_m &\equiv& 3-\Delta_\phi \sim\frac{1}{\sqrt{N}}.
\eea
For example, if we try to estimate the position at which the jump occurs at each $N$, we find excellent fits ($R^2 \gtrsim 99\%$) to $\Delta_{\phi} = 3-a/\sqrt{N}$ behavior with $a \sim 3.5 \pm 1.5$, where the precise value obtained depends on the chosen points, the details of the fit procedure, the inclusion of subleading corrections, etc.\footnote{Coefficients at the lower end of this range are perhaps more likely, both because the jumps will shift to the right at higher $\Lambda$ and because lower values seem to be favored after including subleading $1/N$ corrections in the fit. We also find fits consistent with this range using the locations of the kinks in the singlet bounds. But we defer a more detailed analysis until we have higher-precision data.} By contrast, we find that assuming a $1/N$ scaling generally leads to much poorer fits ($R^2 < 95\%$). It will be interesting in future work to compute this coefficient more precisely. In typical known theories with a proper $SO(N)$ global symmetry, like the critical $O(N)$ vector models, the anomalous dimensions of $SO(N)$ vectors scale as $1/N$ in the large $N$ expansion. Thus, the above scaling behavior seems to be exotic if the $SO(N)$ symmetry is the proper global symmetry of the underlying theory.

In the large $N$ limit, we see that the non-singlet sectors play an important role in the analysis. From the bootstrap point of view, bounds in the non-singlet sectors are typically stronger (lower) at larger $N$ \cite{Kos:2013tga, Iliesiu:2017nrv}. When this is the case they are guaranteed to be finite in the large $N$ limit. The difference between the singlet and non-singlet sectors can be clearly explained in the large $N$ extremal solutions which we have observed to coincide with generalized free field theory.

Generalized free field theories are non-local CFTs that describe the leading behavior of general large $N$ CFTs. In these theories, the 4-point correlator $\langle \phi_i(x_1)\phi_j(x_2)\phi_k(x_3)\phi_l(x_4)\rangle$ of the $SO(N)$ vector scalar $\phi_i$ is obtained through Wick contractions
\bea
\langle \phi_i(x_1)\phi_j(x_2)\phi_k(x_3)\phi_l(x_4)\rangle &&= \frac{1}{x_{12}^{2\Delta_\phi}x_{34}^{2\Delta_\phi}}\times \nn\\
&&\left(\delta_{ij}\delta_{kl} \left(1+\frac{1}{N}u^{\Delta_\phi}+\frac{1}{N}(\frac{u}{v})^{\Delta_\phi}\right) \right. \nn\\
&&\left.+\frac{1}{2}(\delta_{ik}\delta_{jl}+\delta_{il}\delta_{jk}-\frac{2}{N}\delta_{ij}\delta_{kl})
\left(u^{\Delta_\phi}+\left(\frac{u}{v}\right)^{\Delta_\phi}\right)\right.  \nn\\
&&\left.+\frac{1}{2}(\delta_{ik}\delta_{jl}-\delta_{il}\delta_{jk})
\left(u^{\Delta_\phi}-\left(\frac{u}{v}\right)^{\Delta_\phi}\right)\right), \label{GFFON}
\eea
where $x_{ij}=x_i-x_j$ and $(u,v)$ are the standard conformal invariant cross ratios $u= \frac{x_{12}^2x_{34}^2}{x_{13}^2x_{24}^2}$, $v= \frac{x_{14}^2x_{23}^2}{x_{13}^2x_{24}^2}$.
The three terms in the right hand side of (\ref{GFFON})  give contributions from singlet, traceless symmetric and anti-symmetric representations of $SO(N)$ symmetry.
In the $N\rightarrow\infty$ limit the singlet sector becomes trivial, and there is no non-unit singlet operator that can appear in the conformal partial wave decomposition of 4-point correlator $\langle \phi_i(x_1)\phi_j(x_2)\phi_k(x_3)\phi_l(x_4)\rangle$ with nonzero OPE coefficient.
On the other hand, double-trace operators appear in both the symmetric and anti-symmetric sectors of the conformal partial wave decomposition of the 4-point correlator with coefficients of order $O(N^0)$. 

\begin{figure}[!t]
\includegraphics[scale=0.7]{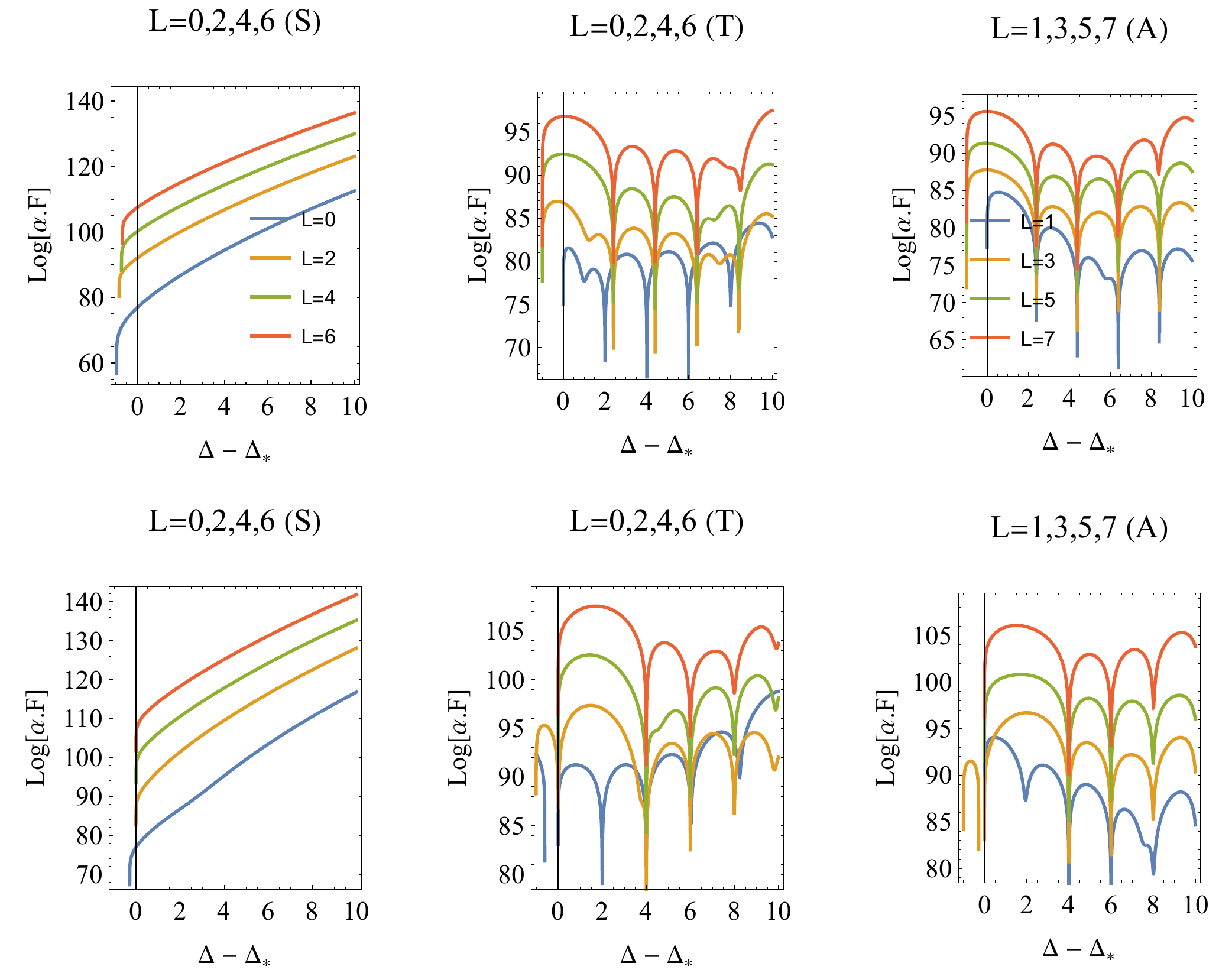}
 \begin{flushright}
\caption{Plots of the extremal functions $\alpha\cdot V_{S/T/A,\Delta_*+x,L}$ in the variable $x=\Delta-\Delta_*$, for each spin $L$, at $\Delta_{\phi}=2.2, \Delta_{T}=4.4$ (first line) and $\Delta_{\phi}=2.9999, \Delta_{T}=5.999805$ (second line), where $\Delta_{T}$ is the scaling dimension of the lowest traceless symmetric scalar appearing in the $\phi_i\times\phi_j$ OPE. The extremal functions are computed at $\Lambda=31$. $S/T/A$ denote singlet/traceless symmetric/anti-symmetric sectors in the $SO(N)$ crossing equation (\ref{ONce}). Here $\Delta_*$ is the unitary bound for spin $L$ operators (for instance, $\Delta_*=1$ for $L=0$ in $S$ sector), except for the $T$ sector at $L=0$, in which case $\Delta_*$ is given by $\Delta_T$. The top three graphs give spectra of generalized free field theory, in which only double-trace operators appear in the extremal functions. In the plots of the second line, both double-trace operators and higher-spin currents appear in the extremal functions. Note that there is a spurious conserved current ($x=0$) in the $L=1$ extremal function at $\Delta_\phi=2.2$ (third graph).
  } \label{EFMinf}
\end{flushright}
\end{figure}

We use the extremal functional method to extract details on the optimal solution of the crossing equation near the jump.
The extremal functions at $\Delta_\phi=2.2/2.9999$ with $N=\infty$ are shown in the first/second line of Figure \ref{EFMinf}. 
The upper bound is close to $\Delta_T\simeq4.4$ for $\Delta_\phi=2.2$ and $\Delta_T\simeq 6$ for $\Delta_\phi=2.9999$. Note that the point $\Delta_\phi=2.9999$ is slightly to the left of $\Delta_\phi=3$ and the corresponding value $\Delta_T\simeq 6$ locates at the bottom of the jump.

At $\Delta_\phi=2.2$, there are no operators in the singlet sector, while in the traceless symmetric and anti-symmetric sectors only double-trace operators appear. This is consistent with the fact that in the region $\Delta_\phi < 3$, the extremal solution is given by generalized free field theory. However, at $\Delta_\phi=2.9999$, we observe an interesting mixing in the spectrum.
Both double-trace operators and a series of conserved higher spin currents (but not a spin $0$ current) appear in the spectrum.\footnote{The readers should be reminded that spurious operators could appear in the extremal spectra and not all the zeros in the extremal function are necessarily identified with physical operators. For instance, in the third graph of Figure \ref{EFMinf} there is a spurious spin $1$ conserved current. These can be distinguished by the fact that their OPE coefficients in the extremal solution are vanishingly small.} As there is a scalar with $\Delta=3$, the higher spin currents are likely to be constructed with fermion bilinears. In a free fermion theory, the spin $0$ current $j_0=\bar{\psi}\psi$ can not appear in the $j_0\times j_0$ OPE due to parity symmetry. This explains the absence of a scalar current in $L=0$ singlet sector. In the large $N$ limit, the fermion bilinear 4-point correlator contains two parts: the disconnected part given by generalized free field theory and a connected part containing contributions from higher spin currents. Besides the generalized free field theory, the connected part of the 4-point correlator also provides a solution to the crossing equation \cite{Turiaci:2018dht}. 

The mixing in the spectrum suggests that the extremal solution at the top of the jump $(\Delta_\phi=3, \Delta_T=8)$ likely corresponds to a linear combination of the generalized free field theory solution and the free fermion solution, such that the $\Delta_T=6$ operator is absent. In fact, by explicit construction one can establish this and also show that the extremal solution at the top of the jump contains a series of higher-spin conserved currents. We will not dwell on this construction or its spectrum, as the detailed solution to the crossing equation and its relation to the free fermion bilinear 4-point correlators \cite{Dolan:2000ut,Maldacena:2011jn, Turiaci:2018dht, Bedhotiya:2015uga, Li:2019twz, Kalloor:2019xjb} has been presented in the parallel work \cite{He:2020azu}.  However, we expect that similar mixing phenomena could be seen numerically at large but finite $N$. In consequence, the original kinks may correspond to certain  underlying theories mixed with a generalized free field theory. The mixing problem can be solved by imposing more constraints in the bootstrap implementation, for example by imposing a finite $c$ central charge or conserved current central charge. Both of the central charges are significantly different between a physical theory at finite $N$ and a generalized free field theory, and therefore they can be used to separate the underlying physical theory from unphysical solutions \cite{AELP}. 
Similar ideas have been used in previous bootstrap studies (e.g.~\cite{Dymarsky:2017xzb, Karateev:2019pvw}) to exclude generalized free field theory solutions.

At large $N$ the kink locations show an interesting behavior which suggests that the putative underlying theory of the kinks may be a deformation from free fermion theory. If they correspond to physical theories, one hopes that they could be studied perturbatively through a $1/N$ expansion and then one may try to compare the bootstrap results with perturbative predictions of certain known Lagrangian theories. This has been done in 3D where the $x$-positions of the kinks are close to the large $N$ results of QED$_3$ \cite{Li:2018lyb}. However, in 4D non-supersymmetric CFTs, like the CBZ fixed points with gauge group $SU(N_c)$, there are two control parameters: the flavor number $N_f$ and the degree of the color group $N_c$. In this two dimensional parameter space, only a very special line of $(N_f, N_c)$ could possibly saturate the bootstrap bounds. One simple possibility is that the kinks pick out a theory at or near the top of the conformal window at a given $N_f$.\footnote{In the Veneziano limit, a theory with $N_f = \frac{11}{2} N_c - n$ has an anomalous dimension $\gamma_m \sim \frac{22}{25} \frac{n}{N_f}$ (see e.g.~\cite{Ryttov:2017kmx}), where in physical theories $n$ is half-integer. The coefficient $\frac{22}{25} n = 0.88 n$ can be compared with our initial estimate $\frac{a}{\sqrt{2}} \sim 1.4 - 3.5$ after matching $N = 2 N_f^2$. It will be interesting to do this comparison with higher precision data.} Establishing this or some other scenario using the bootstrap results will require quite high precision.\footnote{A related issue is that there are many large $N$ equivalences between different fixed points, e.g.~\cite{Bond:2019npq}.} More CFT data from both the bootstrap and perturbative sides will likely be needed to extract a firm conclusion about the large $N$ behavior. However in the next section we will see that interesting comparisons can still be made at small $N$ after inputting the full flavor group of the CBZ fixed points.

\FloatBarrier\section{Bootstrapping fermion bilinears in 4D}
\label{sec:cbz}

In this section we will further explore the connection between the bootstrap kinks and the CBZ fixed points. We'll also study coincidences between bootstrap bounds with different global symmetries, making a connection between the $SO(N)$ vector bounds described above and bounds from 4-point functions of $SU(N_f)_L \times SU(N_f)_R$ bi-fundamentals that would be applicable to the CBZ fixed points.

In one of the simplest versions, the CBZ fixed points contain $N_f$ flavors of massless Dirac fermions in the fundamental representation of a gauge group $SU(N_c)$. For a given $N_c$, the IR fixed point is realized within an interval of $N_f$, the conformal window. Near the upper bound of the conformal window, the theory is weakly coupled and can be studied using perturbation theory. However, the theory becomes strongly coupled near the lower bound of the conformal window and it is extremely difficult to determine the critical flavor number. The theory can be straightforwardly generalized to different gauge groups and representations carrying more color indices. Even in the simplest version of the CBZ fixed points, the CFT landscape is significantly more complicated than it is for conformal QED$_3$: there are two parameters $N_f$ and $N_c$ related to each fixed point, and since  the bootstrap only focuses on gauge-invariant operators, we lose information about both the gauge group and the representations of the fermions. As a result, it is quite subtle to interpret bootstrap results in terms of known gauge theories.

QCD  with gauge group $SU(N_c)$ ($N_c\geqslant3$) and $N_f$ massless fundamental fermions has chiral symmetry $SU(N_f)_L\times SU(N_f)_R$, which is unbroken in  the IR conformal phase. The lowest  gauge-invariant operators are fermion bilinears
\bea
\cO^{\bar{i}}_i\equiv\bar{\psi}_L^{\bar{i}}\psi_{Ri},
\eea
in which $\psi_{R/L}$ are Weyl components of fundamental fermions and the color indices are contracted implicitly.
The theory has parity and charge conjugation symmetry, under which the Weyl components change their chirality. The two flavor groups $SU(N_f)_{L/R}$ are symmetric and will not be distinguished in the bootstrap implementation.
The fermion bilinears transform in the bi-fundamental representation
{\fontdimen8\textfont3=0.8pt
$\overbar{\square}\times\square$ }
of the chiral symmetry and they provide natural candidates for a bootstrap study. Specifically, we can bootstrap the 4-point correlator
\beq
\langle \cO^{\bar{i}}_i(x_1) {\cO^\dagger}^{\bar{j}}_j(x_2)\cO^{\bar{k}}_k(x_3) {\cO^\dagger}^{\bar{l}}_l(x_4)\rangle. \label{4DSUN2}
\eeq

It is straightforward to obtain the crossing equation following the general procedure provided in \cite{Rattazzi:2010yc}.
For each $SU(N_f)$ contained in the chiral symmetry, the representations that can appear in the $\cO\times\cO^\dagger$ or $\cO\times\cO$ OPE are the singlet ($S$), adjoint (Adj), symmetric ($T$), or anti-symmetric ($A$) representations with suitable spin selection rules. The crossing equation of (\ref{4DSUN2}) then includes a symmetrized double copy of the above $SU(N_f)$ representations.
An explicit formula for the crossing equation has been given  in \cite{Nakayama:2016knq}, in which a bootstrap study aimed at the CBZ fixed points was performed. This work resulted in a  lower bound on the scaling dimension $\Delta_{\bar{\psi}\psi}$ of the fermion bilinear $\cO^{\bar{i}}_i$ under the assumption that the presumed IR fixed point can be realized within a given lattice regularization.

The crossing equation can be written in a compact form \cite{Nakayama:2016knq}
\bea
\sum_{O\in\cO\times\cO^\dagger}\lambda^2_O V_{S,S,\Delta,\ell}^{(\pm)}+\sum_{O\in\cO\times\cO^\dagger}\lambda^2_O V_{\text{Adj,Adj},\Delta,\ell}^{(\pm)}+\sum_{O\in\cO\times\cO^\dagger}\lambda^2_O V_{\text{Adj},S,\Delta,\ell}^{(\pm)}+\nn\\
\sum_{O\in\cO\times \cO}\lambda^2_O V_{T,T,\Delta,\ell}^{(+)}+\sum_{O\in\cO\times \cO}\lambda^2_O V_{T,A,\Delta,\ell}^{(-)}+ \sum_{O\in\cO\times \cO}\lambda^2_O V_{A,A,\Delta,\ell}^{(+)}+\cdots=0,  \label{4DSUN2ce}
\eea
where $V_{X}^{\pm}$ are $9$-component vectors. Details on the vectors are presented in Appendix \ref{app:crossing}.
Contributions of the sectors $V_{S,\text{Adj}}$ and $V_{A,T}$ are suppressed in (\ref{4DSUN2ce}) as they will not give new constraints. We also suppressed their complex conjugate representations in the above crossing equation. (For brevity, these representations will all be implicitly assumed in the OPE, crossing equations, and branching rules which we will discuss later.)
We are interested in bounds on the scaling dimensions of the scalars and will mainly focus on the scalars in the $(S,S)$ and $(T,T)$ sectors.

Surprisingly, the bound on the scaling dimension of the lowest scalar in the singlet sector obtained from the $SU(N_f)\times SU(N_f)$ bi-fundamental crossing equation (\ref{4DSUN2ce}) is exactly the same (up to the precision of our binary search) as that from the $SO(N^*)$ vector bootstrap, given $N^*=2N_f^2$! On the other hand, the bound on scaling dimension of the lowest scalar in the $(T,T)$ sector is weaker than that from $SO(N^*)$ vector bootstrap. As we will discuss later, it can be made identical to the latter by imposing some additional conditions.

In the next section we will study the general relations between bootstrap bounds with different global symmetries, which will be important to giving a proper interpretation of the bootstrap results. 

\FloatBarrier\subsection{Coincidences between bootstrap bounds with different global symmetries}

\subsubsection{Coincidence of singlet bounds}
Coincidences between bootstrap bounds on the scaling dimensions of the singlet scalars seem to be quite general. One example is the coincidence between the singlet bounds from the bootstrap with $SO(2N)$ vector and $SU(N)$ fundamental scalars \cite{Poland:2011ey}. Bounds arising from 4-point functions of $SO(N^2-1)$ vector and $SU(N)$ adjoint scalars \cite{Nakayama:2017vdd, Li:2018lyb} are also known to coincide with each other. By comparing the symmetries of external scalars involved in the bound coincidences, one may notice that the representations of different groups that lead to the same bounds actually have the same dimension (or number of components), among which the $SO(N)$ vector realizes the largest symmetry group with a representation of the given dimension. One may expect a more general statement:\footnote{We have so far tested three examples for this conjecture. It would be interesting to check this statement with other continuous symmetry groups. In this work we do not discuss discrete symmetry groups, but one may also wonder if there are similar coincidences for discrete symmetry groups as well. An example is given in  \cite{Stergiou:2018gjj}, which shows that bounds on the singlet scaling dimension from the bootstrap with $SO(N)$ symmetry are the same as bounds assuming a discrete symmetry $C_N=S_N\ltimes \mathbb{Z}_2^N$.}

{\it Given a scalar which forms an $\cal N$-dimensional representation $\cal R$ of a group $\cal G$, the bootstrap bound on the lowest singlet scalar obtained from the 4-point correlator $\langle\cal R \bar{\cal R}\cal R \bar{\cal R}\rangle$ will coincide with the singlet bound from the $SO(\cal N)$ vector bootstrap.}

Besides coincidences between the singlet bounds, we can further ask if the whole spectra of the extremal solutions to the crossing equations, after decomposing the  $SO(\cal N)$ representations into the representations of its subgroup $\cal G$, are also identical with each other. If this is true, then it means the extremal solutions to the crossing equation of $\langle\cal R \bar{\cal R}\cal R \bar{\cal R}\rangle$ at the boundary are actually enhanced to $SO(\cal N)$ global symmetry. We'll use the $SO(2N_f^2)$ vector and $SU(N_f)\times SU(N_f)$ bi-fundamental bootstrap as an example for this study.

\begin{figure}[!htb]
\includegraphics[scale=0.7]{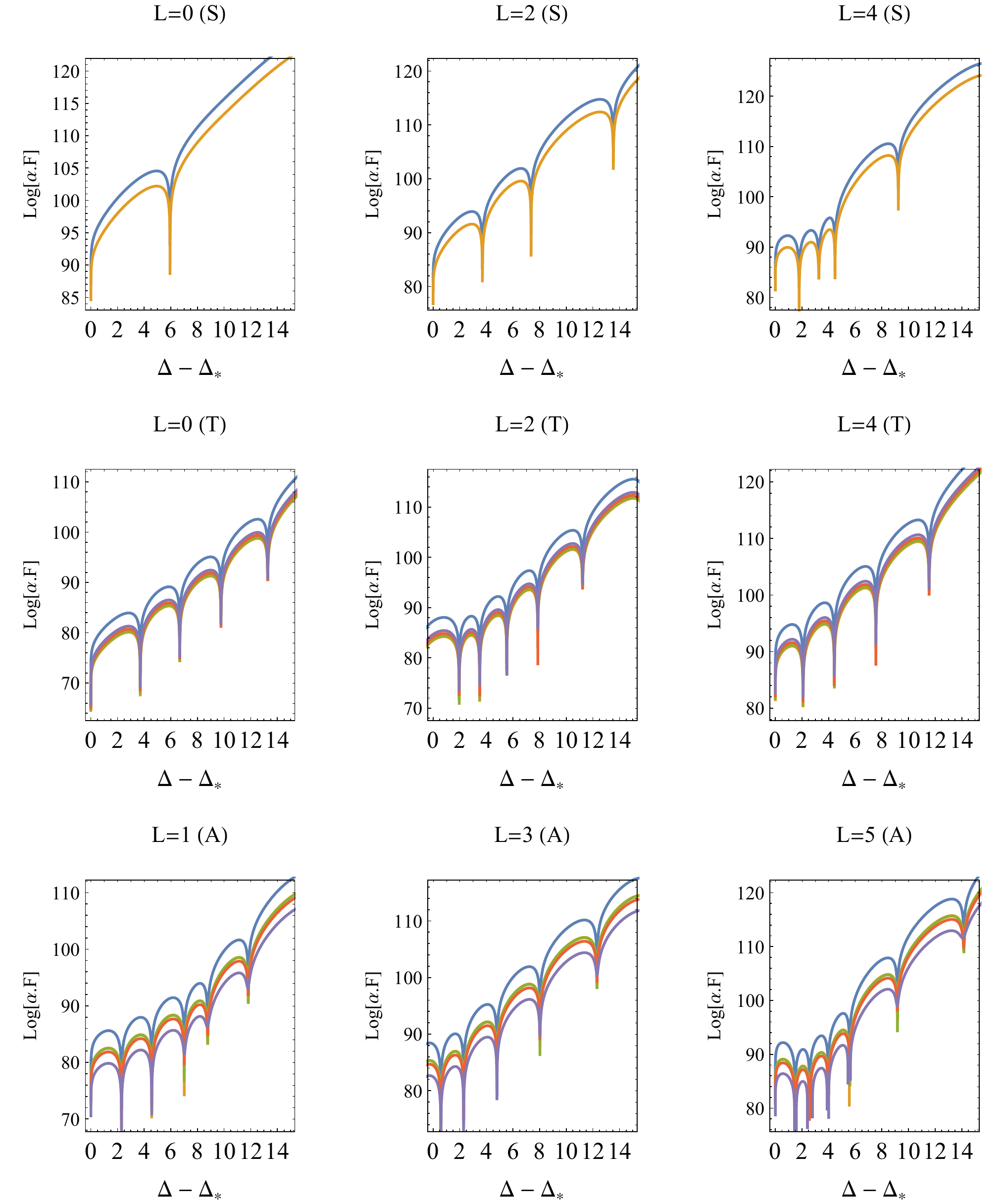} 
\begin{flushright}
\caption{Extremal functions at ($\Delta_{\phi/\cO}=2.0, \Delta_{O_S}\simeq7.04916$), where $O_S$ is the lowest singlet scalar appearing in either the $\phi_i\times\phi_j$ or $\cO^{\bar{i}}_i\times{\cO^\dagger}^{\bar{j}}_j$ OPEs. The extremal functions are computed at $\Lambda=25$. $S/T/A$ denote singlet/traceless symmetric/anti-symmetric sectors in the $SO(N)$ crossing equation (\ref{ONce}). In each graph, the top blue line gives the extremal function from the $SO(32)$ vector bootstrap, while the lower lines with different colors give extremal functions of several sectors in the $SU(4)\times SU(4)$ bi-fundamental bootstrap. In the graphs, the $x$-component is the scaling dimension of the operator in the sector, shifted by $\Delta_*$,  which is $7.04916$ for the $S$ sector with $L=0$, and the unitary bound for all other sectors.
  } \label{EFMmatch}
\end{flushright}
\end{figure}

In Figure \ref{EFMmatch}  we present the extremal functions of the $SO(32)$ vector bootstrap compared to the $SU(4)\times SU(4)$ bi-fundamental bootstrap at $\Delta_{\phi}/\Delta_{\text{bf}}=2$, which is slightly to the left of the kink shown in Figure \ref{4DONs}. The upper bound locates in the range $\Delta_S\in(7.04916, 7.04917)$ at $
\Lambda=25$, slightly weaker than the upper bound  $\sim6.9$ computed in Figure \ref{4DONs} at $\Lambda=31$.
The labels $(L=x, S/T/A)$ above the graphs denote the spin ($L$) and representations (singlet, traceless symmetric and anti-symmetric representations of $SO(32)$) corresponding to the extremal functions.
The top blue line in each graph gives extremal function from the $SO(32)$ vector bootstrap, and the lower lines with different colors give extremal functions of several sectors in the $SU(4)\times SU(4)$ bi-fundamental bootstrap (\ref{4DSUN2ce}).
In particular, the extremal functions in the singlet sector have a one-to-one mapping relation, while the extremal functions in the $T/A$ sectors of $SO(32)$ map to several sectors of $SU(4)\times SU(4)$. 

Following the notation used in the $SU(4)\times SU(4)$ bi-fundamental crossing equation (\ref{4DSUN2ce}), the mapping between the $SO(32)$ and $SU(4)\times SU(4)$ extremal functions is given by\footnote{As before, we suppress the extra sectors $ V_{S, \text{Adj}}^{(\pm)}$, $V_{A, T}^{(-)}$ and the complex conjugate representations. They appear in the decomposition but do not introduce new constraints or extremal functions in the bootstrap.}
\bea
SO(2N_f^2) & \hspace{3cm}  & SU(N_f)\times SU(N_f) \nn \\
S~~ ~&\longleftrightarrow&  ~~~V_{S,S}^{(+)},  \label{branching1}\\
T~~ ~&\longleftrightarrow& ~~~ V_{\text{Adj},\text{Adj}}^{(+)} \simeq V_{\text{Adj},S}^{(+)}\simeq V_{T,T}^{(+)} \simeq V_{A,A}^{(+)}, \label{branching2} \\
A~~~ &\longleftrightarrow& ~~~V_{S,S}^{(-)} \simeq V_{\text{Adj},\text{Adj}}^{(-)} \simeq V_{\text{Adj},S}^{(-)}\simeq V_{T,A}^{(-)}. \label{branching3}
\eea
In Figure \ref{EFMmatch} only the extremal functions of the first three lowest spins in each representation are shown, while similar agreement also appears in the extremal functions with higher spins.

The conclusion of the above analysis is that the upper bound on the scaling dimension of the lowest singlet scalar obtained from the $SU(N_f)\times SU(N_f)$ bi-fundamental bootstrap is given by an extremal solution with fully enhanced $SO(2N_f^2)$ global symmetry.

We have also checked that there is a similar mapping between the spectra of extremal solutions in the $SO(2N)$ vector bootstrap and the $SU(N)$ fundamental bootstrap.\footnote{A similar relation between extremal spectra of the $SO(N^2-1)$ vector bootstrap and $SU(N)$ adjoint bootstrap was also checked in \cite{AELP}.} For the $SU(N)$ fundamental bootstrap, there are four sectors in the crossing equation corresponding to the singlet ($S$), adjoint (Adj), symmetric ($T$) and anti-symmetric ($A$) representations of the $SU(N)$ group:
\beq
\sum_{S}\lambda_\cO^2 V^{(\pm)}_{S,\Delta,\ell}+\sum_{\text{Adj}}\lambda_\cO^2 V^{(\pm)}_{\text{Adj},\Delta,\ell}+\sum_{T}\lambda_\cO^2 V^{(+)}_{T,\Delta,\ell}+\sum_{A}\lambda_\cO^2 V^{(-)}_{A,\Delta,\ell}=0. \label{SUce}
\eeq

The $SU(N)$ crossing equation (\ref{SUce}) was previously presented in \cite{Rattazzi:2010yc}. We'll show the form of the explicit vectors appearing in this crossing equation later in (\ref{SUceM}).  Like the $SU(N_f)\times SU(N_f)$ bi-fundamental bootstrap,
 the extremal spectra coincide across different sectors. Specifically we see the mapping
\bea
SO(2N) & \hspace{3cm}  &  SU(N) \nn \\
S~~ ~&\longleftrightarrow&  ~~~V_{S}^{(+)},  \label{branching4}\\
T~~ ~&\longleftrightarrow& ~~~ V_{\text{Adj}}^{(+)} \simeq V_{T}^{(+)}, \label{branching5}\\
A~~~ &\longleftrightarrow& ~~~ V_{S}^{(-)}\simeq V_{\text{Adj}}^{(-)} \simeq V_{A}^{(-)}. \label{branching6}
\eea

Apparently the mappings (\ref{branching1}-\ref{branching6}) are nothing else but the $SO(\cal N)\rightarrow {\cal G}$ branching rules for $SO(\cal N)$ representations. It is quite amazing that although the crossing equations, such as equations (\ref{ONce}), (\ref{4DSUN2ce}), and (\ref{SUce}), are endowed with different forms, the numerical bootstrap can figure out precise branching rules just from general consistency conditions. Note that a necessary condition for the above branching rules is that the external scalar in the representation $\cal R$ of group $\cal G$ should have the same number of degrees of freedom (or dimension of $\cal R$) as the $SO(\cal N)$ vector, otherwise the extremal solutions from the two different crossing equations cannot contain the same information and a one-to-one mapping between the extremal solutions is not possible. We will come back to this point when discussing a possible approach to avoid such symmetry enhancement.

\subsubsection{Coincidence of non-singlet bounds}

According to the above analysis, the coincidence of the singlet bounds follows the branching rules (\ref{branching1}-\ref{branching6}). In these branching rules the singlet sector is on a similar footing as the non-singlet representations -- the only relevant property is that it relates to a one-to-one mapping. One may expect a similar coincidence between bounds of non-singlet operators as long as the representations appearing in the branching rules are treated carefully.

Let us take the $SO(2N_f^2)$ vector and $SU(N_f)\times SU(N_f)$ bi-fundamental bootstrap for example. In the $SO(2N_f^2)$ vector bootstrap, the non-singlet scalar appears in the traceless symmetric representation. Its branching rule to $SU(N_f)\times SU(N_f)$ is given by (\ref{branching2}). Without extra assumptions in the bootstrap conditions, the bound on the scaling dimension of the lowest  $SO(2N_f^2)$ traceless symmetric scalar is stronger than that of the $SU(N_f)\times SU(N_f)$ symmetric-symmetric scalar. However, if we impose an assumption that all the lowest scalars in the four sectors on the right hand side of  (\ref{branching2}) have the same scaling dimension, then the bound is exactly the same as that of $SO(2N_f^2)$ traceless symmetric scalar. Here the $SO({\cal N})\rightarrow \cal G$ branching rule plays the same role as for the singlet bound.

In general, operators with different representations receive different quantum corrections and it is unlikely for them to have exactly the same scaling dimension without extra symmetries. On the other hand, in the planar limit of a large $\cal N$ theory, the composite operators appear in the OPE are factorized and the above assumption indeed can be satisfied at leading order. In this case, the bound on the $SO(\cal N)$ traceless symmetric scalar could be considered as the leading order result for these representations up to $1/{\cal N}$ corrections.

These results lead to two immediate questions: from the numerical bootstrap point of view, why do we always have such a drastic symmetry enhancement?  And going back to the kinks we found in Figure \ref{4DONs}, assuming these kinks relate to full-fledged theories, are the enhanced $SO(\cal N)$ global symmetries physical or just caused by the bootstrap algorithm? We address these two questions in the following subsections.

\FloatBarrier\subsection{A proof of the coincidences between bootstrap bounds}

The ``symmetry enhancement" phenomena was first observed in \cite{Poland:2011ey}, where the authors discussed relations between singlet bounds with different symmetries. Physically a symmetry enhancement from $SU(N)$ to $SO(2N)$ would be rare, especially away from very small values of $N$. The coincidence in the bootstrap bounds should be ascribed to the bootstrap algorithm and some hidden properties of the representations of symmetries. Here we'll give a proof for the coincidence between the bounds on the singlet scaling dimensions obtained from the $SO(2N)$ vector and $SU(N)$ fundamental bootstrap. Our proof can be straightforwardly generalized to different symmetries. We plan to give a more detailed study of this problem in a follow up work.

First, one can straightforwardly show an inequality between the upper bounds on the scaling dimension of the leading singlet ($\Delta_S$), assuming the existence of either a vector under an $SO(2N)$ symmetry or a fundamental under an $SU(N)$ symmetry:
\beq
\Delta_S|_{SO(2N)} ~~\leqslant ~~\Delta_S|_{SU(N)}. \label{so2su}
\eeq
To see this, let us compare the set of solutions to the crossing equations of 4-point functions of $SO(2N)$ vector ($\cal S_{O}$) or $SU(N)$ fundamental ($\cal S_{U}$) scalars. Since $SU(N)\subset SO(2N)$, we can always decompose solutions to the $SO(2N)$  crossing equation into those of $SU(N)$.
Note in the decomposition of $SO(2N) \rightarrow SU(N)$, as shown in the mapping of Eqs. (\ref{branching4}-\ref{branching6}), there are no new $SU(N)$ singlets appearing besides the original $SO(2N)$ singlet.\footnote{This condition is important, otherwise the definition of the bootstrap problem could be implicitly modified. For instance, in the decomposition $SO(N+1)\rightarrow SO(N)$,  besides the original $SO(N+1)$ singlet ${\cal O}_{S}$, an extra $SO(N)$ singlet appears from the decomposition of the $SO(N+1)$ traceless symmetric scalar, which in principle could have a lower scaling dimension than $\Delta_{S}$. Then the bound on scaling dimension of the lowest $SO(N+1)$ singlet is equivalent to the bound on the scaling dimension of the second lowest $SO(N)$ singlet, which is of course higher (or weaker) than the bound of the lowest $SO(N)$ singlet.}  The lowest $SO(2N)$ singlet is also the lowest $SU(N)$ singlet, and the  $SO(2N)$ solution automatically provides a solution (with the same $\Delta_S$) to the $SU(N)$ crossing equation, which could be extremal or not. Therefore we have
\beq
\cal S_{O}\subseteq \cal S_{U},
\eeq
which implies the relation (\ref{so2su}).

Thus, the $SO(2N)$ singlet bound cannot be weaker (higher) than the $SU(N)$ singlet bound. This conclusion was also obtained in \cite{Poland:2011ey}. The coincidence of the $SO(2N)$ and $SU(N)$ bounds further suggests that the extremal solution of $SO(2N)$ is also extremal for $SU(N)$. Alternatively, any point $(\Delta_\phi, \Delta_S)$ that is excluded by the $SO(2N)$ vector bootstrap is also excluded by the $SU(N)$ fundamental bootstrap.

We will study the structure of the crossing equation. For this purpose, it is helpful to write the crossing equations (\ref{ONce}) and (\ref{SUce}) into matrix forms:
\bea
M_{SO(2N)}=\left(
\begin{array}{ccc}
 0 & F & -F \\
 F &  \left(1-\frac{1}{N}\right)F & F \\
 H & -\left(\frac{1}{N}+1\right)H & -H \\
\end{array}
\right),  
\eea
\bea
M_{SU(N)}=\left(
\begin{array}{cccccc}
 0 & 0 & F & -F & F & -F \\
 0 & 0 & H & -H & -H & H \\
 F & F & F \left(1-\frac{1}{N}\right) & F \left(1-\frac{1}{N}\right) & 0 & 0 \\
 H & H & H \left(-\left(\frac{1}{N}+1\right)\right) & H \left(-\left(\frac{1}{N}+1\right)\right) & 0 & 0 \\
 F & -F & -\frac{F}{N} & \frac{F}{N} & F & F \\
 H & -H & -\frac{H}{N} & \frac{H}{N} & -H & -H \\
\end{array}
\right), \label{SUceM}
\eea
where the columns are given by the vectors $$(V_{S},~V_{T},~V_{A})$$ and $$({V_S^{(+)}}, ~ {V_S^{(-)}}, ~ {V_{\text{Adj}}^{(+)}}, ~ {V_{\text{Adj}}^{(-)}}, ~  {V_{T}^{(+)}}, ~ {V_{A}^{(-)}})$$ which appear in the crossing equations (\ref{ONce}) and (\ref{SUce}) respectively.

In the $SO(2N)$ vector bootstrap, the bootstrap problem can be rewritten in the form
\bea
\left(
 \alpha _1 ~ \alpha _2 ~ \alpha _3 
\right) \cdot  \left(
\begin{array}{ccc}
 0 & F & -F \\
 F &  \left(1-\frac{1}{N}\right)F & F \\
 H & -\left(\frac{1}{N}+1\right)H & -H \\
\end{array}
\right) &=&  (\alpha_S ~ \alpha_T ~\alpha_A)    \succcurlyeq (0~0~0),  \nn\\ 
&&~~~\forall \Delta \geqslant \Delta_S \text{ or unitary bound.~~} \label{onmatrix}
\eea
The bootstrap algorithm is then to test if such linear functionals $\alpha_i$ exist for a given $(\Delta_\phi, \Delta_S)$. 

Similarly, we can rewrite the $SU(N)$ fundamental bootstrap problem as
\bea
\left(\beta _1~\beta _2 ~ \beta _3 ~ \beta _4 ~ \beta _5 ~ \beta _6\right) \cdot\left(
\begin{array}{cccccc}
 0 & 0 & F & -F & F & -F \\
 0 & 0 & H & -H & -H & H \\
 F & F &  \left(1-\frac{1}{N}\right) F&  \left(1-\frac{1}{N}\right)F & 0 & 0 \\
 H & H &  -\left(\frac{1}{N}+1\right) H & -\left(\frac{1}{N}+1\right)H & 0 & 0 \\
 F & -F & -\frac{F}{N} & \frac{F}{N} & F & F \\
 H & -H & -\frac{H}{N} & \frac{H}{N} & -H & -H \\
\end{array}
\right)   ~~~~ ~~~~~~~ \nn\\
=(\beta_{V_S^+} ~ \beta_{V_S^-} ~ \beta_{V_{\text{Adj}}^+} ~ \beta_{V_{\text{Adj}}^-} ~  \beta_{V_{T}^+} ~ \beta_{V_{A}^-})
\succcurlyeq 0_{1\times6} , ~~~  \forall \Delta \geqslant \Delta_S \text{ or unitary bound.} \label{sumatrix}
\eea

Inspired by the branching rule of spectra in the extremal solutions (\ref{branching4}-\ref{branching6}), we want to ask the following question: {\it For a specific choice of $(\Delta_{\phi}, \Delta_S)$, suppose that we have obtained linear functions $\alpha_{S/T/A} $ satisfying the positivity condition (\ref{onmatrix}). Is it possible to construct the linear functions $\beta_{V_X^{\pm}}$ satisfying the positivity conditions (\ref{sumatrix})?} 

In particular, we expect a mapping of the following form between the functions in each sector of the crossing equation:
\bea
\beta_{V_S^+} &=& \alpha_S,   \\
\beta_{V_{\text{Adj}}^+} &=& x_2 \alpha_T,  ~~  \beta_{V_{T}^+}= x_4 \alpha_T,\\
\beta_{V_S^-} &=& x_1\alpha_A,  ~~ \beta_{V_{\text{Adj}}^-} = x_3 \alpha_{A}, ~~ \beta_{V_{A}^-}= x_5 \alpha_A,
\eea
or in a vector form:
\bea
(\beta_{V_S^+} ~ \beta_{V_S^-} ~ \beta_{V_{\text{Adj}}^+} ~ \beta_{V_{\text{Adj}}^-} ~  \beta_{V_{T}^+} ~ \beta_{V_{A}^-})& =& 
(\begin{array}{cccccc}
 \alpha _S & x_1 \alpha _A & x_2 \alpha _T & x_3 \alpha _A & x_4 \alpha _T & x_5 \alpha _A \\
\end{array} \label{eq1}
).
\eea
Here we set the coefficient of the first component $\alpha_S$ to be 1 as a normalization condition.\footnote{In the numerical bootstrap, we typically choose an operator, like the unit operator, to set the overall normalization of our functional. In the extremal solution, the positive value used in this normalization is numerically exponentially small compared to the action of the functional on other operators. Our choice $\beta_{V_S^+} = \alpha_S$ stipulates that if the functional is normalized to 1 on the $SO(2N)$ unit operator, then it will also be normalized to 1 on the $SU(N)$ unit operator.}
Since the functions $\beta_{V_X^{\pm}}$ and $\alpha_{S/T/A}$ are obtained from linear functionals $\alpha_i/\beta_i$ applied to two independent bases of functions $F(u,v)/H(u,v)$, the above equation essentially amounts to a reconstruction of the linear functionals $\beta_i$ for the $SU(N)$ fundamental bootstrap from the linear functionals $\alpha_i$ for the $SO(2N)$ vector bootstrap. 

Using the matrix form of the $SO(2N)$ crossing equation (\ref{onmatrix}), the RHS of (\ref{eq1}) is
\bea
\left(
 \alpha _S ~ x_1 \alpha _A ~ x_2 \alpha _T ~ x_3 \alpha _A ~ x_4 \alpha _T ~ x_5 \alpha _A 
\right) &=& \nn\\
&&\hspace{-6cm}  \left(
 \alpha _1 ~ \alpha _2 ~ \alpha _3 
\right)  \cdot 
 \left(
\begin{array}{cccccc}
 0 &  -x_1 F &  x_2 F&  -x_3 F &  x_4F & -x_5F \\
 F & x_1 F &  \left(1-\frac{1}{N}\right) x_2F &  x_3F &  \left(1-\frac{1}{N}\right) x_4F &  x_5F \\
 H &  -x_1H & -\left(\frac{1}{N}+1\right) x_2H & -x_3H & -\left(\frac{1}{N}+1\right) x_4H & -x_5H \\
\end{array}
\right).
\eea
Then combined with (\ref{sumatrix}) we obtain the conditions
\bea
&&\left(\beta _1~\beta _2 ~ \beta _3 ~ \beta _4 ~ \beta _5 ~ \beta _6\right) \cdot\left(
\begin{array}{cccccc}
 0 & 0 & F & -F & F & -F \\
 0 & 0 & H & -H & -H & H \\
 F & F &  \left(1-\frac{1}{N}\right) F&  \left(1-\frac{1}{N}\right)F & 0 & 0 \\
 H & H &  -\left(1+\frac{1}{N}\right) H & -\left(1+\frac{1}{N}\right)H & 0 & 0 \\
 F & -F & -\frac{1}{N}F & \frac{1}{N}F & F & F \\
 H & -H & -\frac{1}{N} H & \frac{1}{N}H & -H & -H \\
\end{array}
\right) =  ~~~~  \nn\\ 
&&\hspace{1cm}  \left(
 \alpha _1 ~ \alpha _2 ~ \alpha _3 \right)  \cdot 
 \left(
\begin{array}{cccccc}
 0 &  -x_1 F &  x_2 F&  -x_3 F &  x_4F & -x_5F \\
 F &  x_1 F &  \left(1-\frac{1}{N}\right) x_2F &  x_3F &  \left(1-\frac{1}{N}\right) x_4F &  x_5F \\
 H &  -x_1H & -\left(1+\frac{1}{N}\right) x_2H & -x_3H & -\left(1+\frac{1}{N}\right) x_4H & -x_5H \\
\end{array}
\right).~~~~ \label{eq2}
\eea

We will now solve for the linear functionals $\beta_i$ in terms of the $\alpha_i$ using the above equations. 
 Let us denote 
\beq
M_{SU(N)}=\text{diag}\{F,~H,~F,~H,~F,~H\} \cdot  {\cal{M}}_{SU(N)}.
\eeq
We then propose the following ansatz for the solution of (\ref{eq2}):
\beq 
\left(\beta _1  ~ \beta _2    ~ \beta _3  ~ \beta _4    ~ \beta _5  ~ \beta _6 \right)= \left(
 \alpha _1 ~ \alpha _2 ~ \alpha _3 \right)  \cdot {\cal T}_{3\times 6}, \label{Trans}
\eeq
where $\mathcal{T}$ is a  $3\times6$ matrix with $5$ unknown parameters $x_i$. 
The general formula for the transformation matrix $\cal T$ can be obtained by  multiplying ${\cal M}_{SU(N)}^{-1}$ on both sides of equation (\ref{eq2}).
Note that for the single correlator bootstrap, as proved in \cite{Rattazzi:2010yc}, the number of bootstrap constraints is always equal to the number of unknown functions, so in general the matrix of crossing equation is always a square matrix (as in $M_{SO(2N)}$ and $M_{SU(N)}$). The matrices like ${\cal M}_{SU(N)}$ are also non-degenerate so are in general invertible.

Both $\alpha_i$ and $\beta_i$ in (\ref{Trans}) are linear functionals which act on the functions $F$ or $H$. As $F$ and $H$ have different parity symmetry under the transformation $x_2\leftrightarrow x_4$, we expect the linear functionals acting on $F$ or $H$ do not mix with each other. Specifically, $\beta_1, \beta_3, \beta_5$, and $\alpha_1, \alpha_2$ are applied to $F$, and so $\beta_{1,3,5}$ should depend on $\alpha_{1,2}$ only while being independent of $\alpha_3$. Similarly $\beta_{2,4,6}$ act on $H$ and should only depend on $\alpha_3$. Then we have the following constraints on the 
linear transformation $\mathcal T$:
\bea
{\cal T}_{3,1}&=&{\cal T}_{3,3}={\cal T}_{3,5}=0, \nn \\
{\cal T}_{1,2}&=&{\cal T}_{1,4}={\cal T}_{1,6}={\cal T}_{2,2}={\cal T}_{2,4}={\cal T}_{2,6}=0.
\eea
Surprisingly, the above 9 equations can be solved for the 5 variables $x_i$:
\beq
(x_1,~x_2,~x_3,~x_4,~x_5)=\left(\frac{1}{2 N-1},\frac{N-1}{2 N-1},\frac{N^2-1}{N(2 N-1)},\frac{N}{2 N-1},\frac{N-1}{2 N-1}\right), \label{xs}
\eeq
and the transformation matrix $\cal T$ has a unique solution (with the normalization adopted in (\ref{eq1})):
\beq
{\cal T}=
\left(
\begin{array}{cccccc}
 1 & 0 & \frac{1}{1-2 N} & 0 & \frac{1}{2 N-1} & 0 \\
 0 & 0 & \frac{1}{2 N-1}+1 & 0 & \frac{1}{1-2 N}+1 & 0 \\
 0 & \frac{2}{2 N-1} & 0 & \frac{1}{1-2 N}+1 & 0 & \frac{1}{2 N-1}+1 \\
\end{array} 
\right). \label{Ts}
\eeq

A crucial property of the solution for the $x_i$ in (\ref{xs}) is that all these parameters are positive for $N\geqslant 2$. Because of this, as long as the functionals of the $SO(2N)$ crossing equation $\alpha_i$ satisfy the positivity conditions (\ref{onmatrix}), all the functionals $\beta_i$ satisfy the positivity conditions (\ref{sumatrix}) as well. Therefore the linear functionals $\beta_i$ constructed from the linear functionals $\alpha_i$ through (\ref{Trans}) can be used in the $SU(N)$ fundamental crossing equation to exclude the CFT data $(\Delta_{\Phi}, \Delta_S)$. Here it is important that the $SU(N)$ singlet is only decomposed from the singlet of $SO(2N)$, therefore the dimensions on which positivity is imposed,
$$\Delta\geqslant \Delta_S \text{ or unitary bound},$$ 
are precisely the same for both the $SO(2N)$ vector bootstrap and $SU(N)$ fundamental bootstrap. 

This, however, is not true for the non-singlet operators, like the $SO(2N)$ traceless symmetric scalar. The $SO(2N)$ traceless symmetric scalar decomposes into an  $SU(N)$ adjoint scalar and an $SU(N)$ symmetric scalar as in (\ref{branching5}). Therefore the positivity conditions of the $SO(2N)$ vector bootstrap, viewed from its $SU(N)$ decomposed representations, are different from those that would be typically used in the pure $SU(N)$ fundamental bootstrap. In consequence, there is no coincidence in the bootstrap bound on the scaling dimension of the symmetric scalar, unless we impose an extra assumption in the $SU(N)$ fundamental bootstrap that the scaling dimensions of the lowest $SU(N)$ adjoint scalar and $SU(N)$ symmetric scalar are identical.

To summarize, the above computations have shown that, if an assumption on the CFT data $(\Delta_{\phi}, \Delta_S)$ in the $SO(2N)$ vector crossing equation is excluded, i.e., the bootstrap algorithm finds linear functionals $\alpha_i$ which satisfy the positivity condition (\ref{onmatrix}), then such linear functionals can be used to construct linear functionals $\beta_i$ through the transformation (\ref{Trans}). Following this fact, we conclude that any point  $(\Delta_{\phi}, \Delta_S)$ that is excluded by the $SO(2N)$ vector bootstrap is also excluded by the $SU(N)$ fundamental bootstrap, and we obtain another relation between the upper bounds on the scaling dimension of the singlet $\Delta_S$ in the $SO(2N)$ vector bootstrap compared to the $SU(N)$ fundamental bootstrap: 
\beq
\Delta_S|_{SO(2N)} ~~\geqslant ~~\Delta_S|_{SU(N)}. \label{su2so}
\eeq
Combining this with the relation (\ref{so2su}) coming from symmetry arguments,\footnote{This relation can also be proved in a way similar to (\ref{su2so}): given linear functionals $\beta_i$ satisfying positivity conditions (\ref{sumatrix}), one can construct linear functionals $\alpha_i$ satisfying the positivity conditions (\ref{onmatrix}). This is straightforward to show using equation (\ref{eq2}) following the branching rules (\ref{branching4}-\ref{branching6}). We thank Slava Rychkov for suggesting this approach.}
 we find 
\beq
\Delta_S|_{SO(2N)} ~~= ~~\Delta_S|_{SU(N)}.
\eeq

Transformation matrices $\cal T$ between linear functionals similar to (\ref{Ts}) can be computed for crossing equations with more general global symmetries and representations. In particular, they can also be constructed to relate the linear functionals in the $SO(N^2-1)$ vector bootstrap to the $SU(N)$ adjoint bootstrap, those of the $SO(2N_f^2)$ vector bootstrap to the $SU(N_f) \times SU(N_f)$ bi-fundamental bootstrap, and those of the $SO(N)$ vector bootstrap to the bootstrap with a discrete symmetry $C_N$. These computations reveal interesting connections between group representation theory and the conformal bootstrap constraints, and we hope to give a more general argument about when such a construction is possible and has the required positivity conditions in a follow up work.

Physically, a continuous global symmetry corresponds to the existence of a conserved current. One may wonder if the $SO(\cal N)$ symmetry enhancement can be avoided by bootstrapping mixed correlators containing both a scalar $\cal R$ in a representation of dimension $\cal N$ and a conserved current $\cal J_G^{\text{adj}}$ of the global symmetry $\cal G$:
\bea
\langle \cal R\bar{R}R\bar{R}\rangle, ~~~ \langle \cal J_G^{\text{adj}}J_G^{\text{adj}}J_G^{\text{adj}}J_G^{\text{adj}}\rangle,  ~~~\langle \cal R\bar{R}J_G^{\text{adj}}J_G^{\text{adj}}\rangle, ~~~\cdots. \label{subceq}
\eea
The conserved current ${\cal J}_{SO(\cal N)}^A$ of $SO(\cal N)$ symmetry decomposes into several representations of the group $\cal G$
\beq
{\cal J}_{SO(\cal N)}^A \rightarrow \cal J_G^{\text{adj}}+ \cal J_G^{\cal R'},
\eeq
in which the $\cal J_G^{\cal R'}$ are also conserved currents in addition to $ \cal J_G^{\text{adj}}$. Such currents are expected to be absent in a $\cal G$-symmetric theory unless there is a real symmetry enhancement ${\cal G} \rightarrow SO(\cal N)$. 

In the crossing equations of the correlators (\ref{subceq}), we fix the global symmetry to be $\cal G$ and only $ \cal J_G^{\text{adj}}$ appears as an external operator in the crossing equation. Therefore comparing with the $SO(\cal N)$ symmetric mixed correlators (containing a scalar $\phi_i$ and current ${\cal J}_{SO(\cal N)}^A$),
\bea
\langle \phi_i\phi_j\phi_k\phi_l\rangle, ~~~ \langle{\cal J}_{SO(\cal N)}^A {\cal J}_{SO(\cal N)}^A{\cal J}_{SO(\cal N)}^A{\cal J}_{SO(\cal N)}^A\rangle,  ~~~\langle \phi_i\phi_j{\cal J}_{SO(\cal N)}^A{\cal J}_{SO(\cal N)}^A\rangle, ~~~\cdots,   \label{onceq1}
\eea
the mixed correlators with $\cal G$ symmetry (\ref{subceq}) have fewer external degrees of freedom and their extremal solutions cannot be organized into whole representations of the $SO(\cal N)$ symmetry. On the other hand, one may decompose the extremal solutions of (\ref{onceq1}) to the $\cal G$ symmetric representations, and truncate them to a subset which can appear in the crossing equations of (\ref{subceq}).  Now the question is if the truncated part of the extremal solutions to the  $SO(\cal N)$ symmetric scalar-current mixed correlators (\ref{onceq1}) can also provide extremal solutions to the mixed correlators  (\ref{subceq}) with $\cal G$ symmetry. Here the conserved current in the mixed correlator system  (\ref{subceq}) is special, as we can apply conservation conditions/Ward identities to the correlators containing $\cal J_G^{\text{adj}}$, and also impose extra assumptions on the correlators of $\cal J_G^{\text{adj}}$ and on the sectors containing $\cal J_G^{\cal R'}$ to prevent unwanted conserved currents. It would be interesting in future numerical studies to see under what conditions the extremal solutions to the crossing equations from (\ref{subceq}) possess proper $\cal G$ symmetry rather than an enhanced $SO(\cal N)$ symmetry.

\FloatBarrier\subsection{Bounds and kinks after breaking $SO(N)$ symmetry} \label{brkON}
It is not plausible that there will be a physical symmetry enhancement from $SU(N_f)\times SU(N_f)$ to $SO(2N_f^2)$ for 4D CFTs and the coincidence between their singlet bounds can be ascribed to the conformal bootstrap algorithm as described above.
A crucial question then is what is the intrinsic global symmetry of the putative full-fledged theories which may correspond to these kinks? 

 A relevant property of the kinks has been shown in Figure \ref{4DONT}: the anomalous dimension of the $SO(N)$ vector scales as $1/\sqrt{N}$ at large $N$, which is hard to explain if the theory is $SO(N)$ symmetric with a scalar in the fundamental representation. This puzzle could be resolved if the underlying theories have symmetry ${\cal G} \subset
SO(N)$, with rank of order $N_f \sim \sqrt{ N}$. These then can potentially relate to the kinks in the $SO(N)$ vector bootstrap bound through a symmetry enhancement of the $\cal G$-symmetric bootstrap solution to an $SO(N)$-symmetric bootstrap solution which doesn't significantly alter the kink location.

One can further probe this question by asking how the kinks behave when one relaxes the symmetry assumptions:
if the $SO(2N_f^2)$ global symmetry is intrinsic to the theories at the kinks, then by breaking the $SO(2N_f^2)$ symmetry explicitly in the bootstrap setup, the kinks will likely become weaker or even disappear; otherwise it is more likely that the kinks are essentially endowed with a global symmetry  ${\cal G}\subset SO(2N_f^2)$. In the following we'll show that the second possibility is favored by our numerical results.

Our strategy is to impose gaps $\delta_X$ in the bootstrap conditions. We require the lowest operators $\cO$ in certain sectors $V_X$ of the  crossing equation (\ref{4DSUN2ce})  have scaling dimensions $\Delta_\cO$:
$$\Delta_\cO\geq \Delta_*+\delta_X, ~~~~~~~~ \forall ~\cO \in V_X,$$
 where $\Delta_*$ is the unitary bound of operators in $V_X$.
 Without imposing any additional constraints in the bootstrap setup ($\delta_X=0$), the singlet upper bound shows exact $SO(2N_f^2)$ symmetry, and the spectra are replicated in different sectors following the branching rules (\ref{branching1}-\ref{branching3}). Such symmetry enhancement can not be realized if $\delta_X$ break the $SO(2N_f^2)$ symmetry explicitly.

\begin{figure}[!htb]
\includegraphics[scale=0.7]{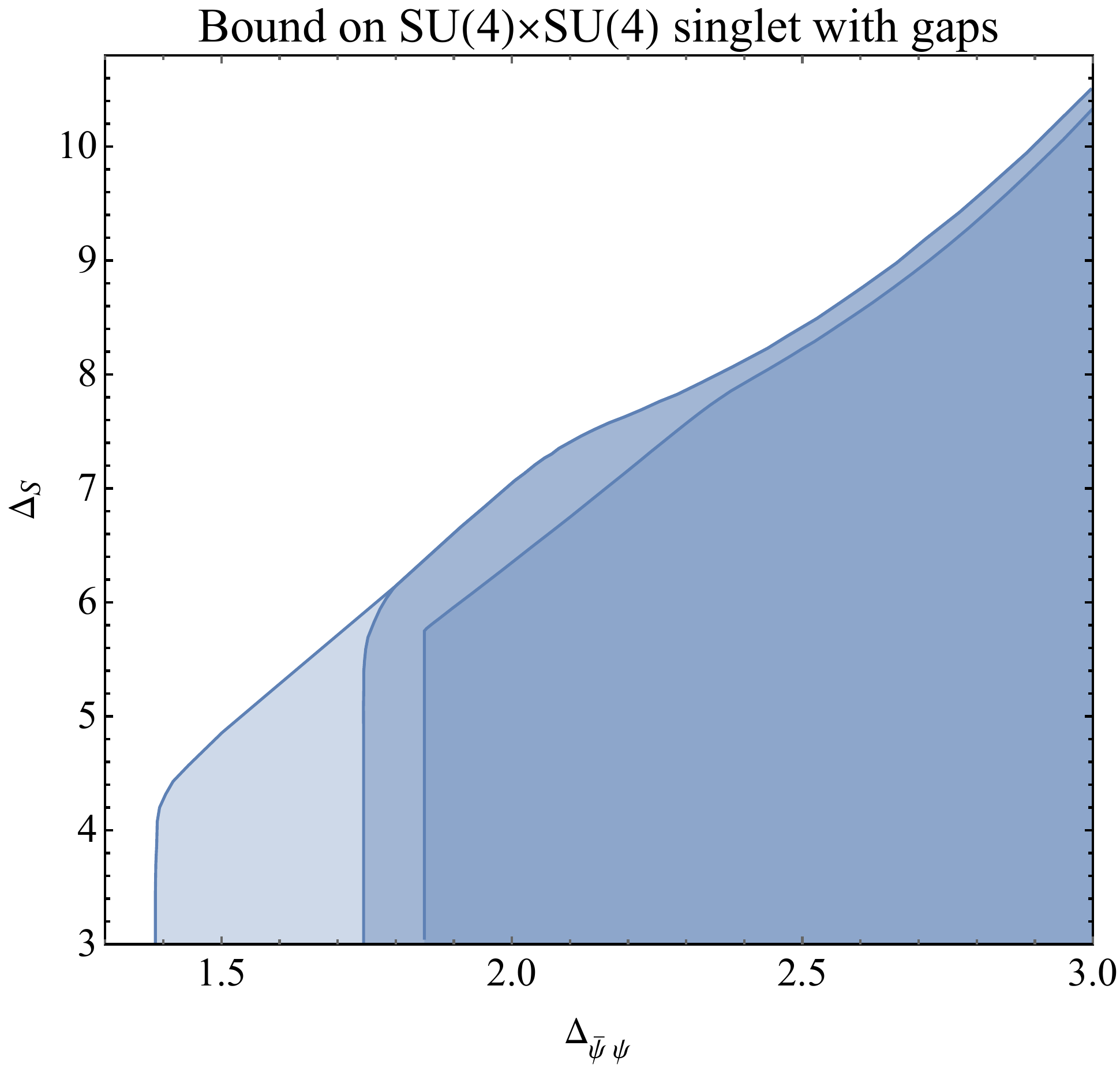}
 \begin{flushright}
\caption{Bounds on the $SU(4)\times SU(4)$ singlet assuming the gaps (\ref{gap3}) and (\ref{gap4}). The bounds are computed at $\Lambda=25$. By imposing the gaps (\ref{gap3}), the bound shows a sharp cut near $\Delta_{\bar{\psi}\psi}=1.38$ (upper line), while the cut moves to $\Delta_{\bar{\psi}\psi}=1.75$ for the gaps (\ref{gap4}) (middle line). The remaining parts of the bounds are only slightly modified by the gaps. A bound on the singlet scaling dimension obtained from the $SO(32)$ vector bootstrap with the gaps (\ref{gapSO}) is also shown in the lowest line. 
  } \label{SU(4)gaps} 
\end{flushright}
\end{figure}

Let us consider the $SU(4)\times SU(4)$ bi-fundamental bootstrap as an example. The results are shown in  Figure \ref{SU(4)gaps}.
We test two sets of gaps in the scaling dimensions of the lowest scalars in the $V_{TT}^{(+)}$ (symmetric-symmetric) sector ($\Delta_{TT,0}$), the $V_{AA}^{(+)}$ (anti-symmetric-anti-symmetric) sector ($\Delta_{AA,0}$), and the leading spin $1$ operator in the $V_{SS}$ (singlet-singlet) sector ($\Delta_{SS,1}$),
\bea
&&\Delta_{TT,0}\geqslant 3, ~~ \Delta_{AA,0}\geqslant 3, ~~\Delta_{SS,1}\geqslant 3.5, \label{gap3}\\
&&\Delta_{TT,0}\geqslant 4, ~~ \Delta_{AA,0}\geqslant 4, ~~\Delta_{SS,1}\geqslant 4. \label{gap4}
\eea
For comparison, in Figure \ref{SU(4)gaps} we also present the $SO(32)$ vector bootstrap result with gaps in the scaling dimensions $\Delta_{T,0}$ of the lowest scalar in the $T$ (traceless symmetric)  sector and in the scaling dimensions $\Delta_{A,1}$ of the the leading spin-1 operator in the $A$ (anti-symmetric) sector:
\beq
\Delta_{T,0}\geqslant 4, ~~ \Delta_{A,1} \geqslant 4. \label{gapSO}
\eeq
The $SO(32)$ symmetry is broken by the gaps (\ref{gap3}) and (\ref{gap4}): the symmetry relating the four sectors $V_{\text{Adj, Adj}}^{(+)}$, $V_{\text{Adj},S}^{(+)}$, $V_{T,T}^{(+)}$ and $V_{A,A}^{(+)}$ is broken and operators in these sectors cannot be unified in representations of $SO(32)$ symmetry. More importantly, the $SO(32)$ conserved current can never be formed when there is a gap in the $V_{S,S}^{(-)}$ sector at spin $1$.

We also want to study putative relations between the kink and fixed points of 4D gauge theories.
Operators carrying multiple flavor indices with even parity are composite operators made from fermion pairs which are gauge invariant. The lowest scalars are made from four fermions and have scaling dimension $\Delta=6$ in the free theory limit. They could receive large anomalous dimensions due to strong interactions but generically these four-fermion operators are expected to be irrelevant $\Delta\geqslant4$. In particular, scalars in the $V_{T,T}^{(+)}$ and $V_{A,A}^{(+)}$ sectors are singlets under the the diagonal subgroup of the flavor symmetry $SU(N_f)_V \subset SU(N_f)\times SU(N_f)$. In certain lattice simulation of the CBZ fixed point, the chiral symmetry is broken by the fermion mass term and only the symmetry $SU(N_f)_V$ is preserved in the regularization. If there are relevant scalars in the $V_{T,T}^{(+)}$ or $V_{A,A}^{(+)}$ sectors, it requires fine-tuning to reach the IR fixed point.

By imposing a gap in the scaling dimensions of the spin 1 operators in the $V_{S,S}^{(-)}$ sector, we impose that the fermion bilinears carry no charge under any additional global symmetries. This is true for the vectorial $U(1)_V$, which is part of the global symmetry of the CBZ fixed point since the meson operator is invariant under $U(1)_V$. The axial counterpart $U(1)_A$ is anomalous. To summarize, it is expected there are no conserved spin $1$ currents in the $V_{S,S}^{(-)}$ sector and a mild gap is expected. Nevertheless, we are not aware of evidence on how large the physical gap could be. We adopt the gaps given in (\ref{gap3}) and (\ref{gap4}) for numerical tests.

 Actually by imposing the gaps (\ref{gap3}) or (\ref{gap4}), the singlet upper bound near the kink becomes slightly stronger, however, the change is not easy to detect in Figure \ref{SU(4)gaps}. One can also check that outside of the sectors where the gaps are imposed, spectra in the extremal functions do not change significantly.
The most significant effects of the gaps are to create the sharp cuts in the left region of the graph. The position of the cut depends on the gaps that are imposed. 
We have checked that to the right of the cut, the extremal solutions do break the $SO(32)$ symmetry:
there are conserved spin 1 operators in the sectors $V_{\text{Adj},\text{Adj}}^{(-)}, V_{\text{Adj},S}^{(-)}, V_{T,A}^{(-)}$,
while not in $V_{S,S}^{(-)}$ due to the gaps in (\ref{gap3}) and (\ref{gap4}).

The behavior of the bounds with different gaps is reminiscent of the bounds obtained from the 3D $U(1)_T$ monopole bootstrap \cite{Chester:2016wrc}. The results suggest that near the kink, most of the operators appearing in the OPE of $SU(N_f)\times SU(N_f)$ bi-fundamental scalars are irrelevant, which is consistent with the  general expectation of gauged fermionic theories. 
It would be interesting to know which gaps will cause the cut to approach the kink, as well as whether there exist any gaps that could cause a cut on the right and help to isolate the kink into a closed region. We leave this problem for future study.  

In the region with larger dimensions, the bound, including the kink (though its position changes slightly), is quite insensitive to the gaps that break $SO(2N_f^2)$ symmetry! This can be viewed as a signal that $SO(32)$ is not the intrinsic symmetry of the theory underlying the kink, even though it appears in the $SO(32)$ bound.  
 One may doubt that the reason the kink is not sensitive to the gaps (\ref{gap4}) might be simply because these operators do not play important roles in the original $SO(32)$ symmetric solution at all. For instance the kink could be a non-local solution to the $SO(32)$ crossing equation without a spin 1 conserved current. To answer this question, we test the effect of imposing the gaps  (\ref{gapSO}) in the $SO(32)$ vector bootstrap. The result is shown in Figure \ref{SU(4)gaps}. By imposing the gaps (\ref{gapSO}), the original kink is excluded and the upper bound changes significantly, which suggests that operators below the gaps (\ref{gapSO}) are necessary to construct the $SO(32)$ symmetric solution at the kink.

It is also possible that the kink remains by breaking the $SO(2N_f^2)$ symmetry to other subgroups like $SU(N_f^2)$ or $SU(N_f')$ with an adjoint representation as the external operator. It is hard to uniquely fix the proper global symmetry of the theory at the kink using the current setup. One may try to constrain the bootstrap results to specifically relate to $SU(N_f)\times SU(N_f)$ symmetry by using mixed correlators consisting of fermion bilinears and $SU(N_f)\times SU(N_f)$ conserved currents (along with gaps forbidding additional symmetry enhancement). By bootstrapping these mixed correlators we may obtain bootstrap results without ambiguities in the symmetry. However, the crossing equations of these mixed correlators involve complicated flavor and spinning indices and a bootstrap study would be quite intricate. We expect that lessons learned in previous work on the conserved current bootstrap \cite{Dymarsky:2017xzb, Dymarsky:2017yzx, Reehorst:2019pzi}, as well as recent improvements in numerical bootstrap algorithms \cite{Go:2019lke, Landry:2019qug, Chester:2019ifh, Erramilli:2019njx}, will all be helpful for this study.

\FloatBarrier\subsection{Bounds near the critical flavor number}
With a large flavor number bounds on the scaling dimensions of singlet scalars show prominent kinks. The kink becomes weaker and seems to disappear near a critical flavor number $N^*$. If there are full-fledged theories underlying these kinks it appears plausible that they correspond to deformations of free fermion theory. In 4D, most known non-supersymmetric CFTs correspond to fixed points of gauge theories. However, in the bootstrap approach we focus on gauge-invariant operators and do not have direct control over the gauge interactions, making it generally difficult to decode the precise underlying theory of the kinks. On the other hand, near the critical flavor number $N^*$, the putative theories may reach certain simplifying limits and some interesting properties could be revealed from the bootstrap results. 

\begin{figure}[!htb]
\includegraphics[scale=0.65]{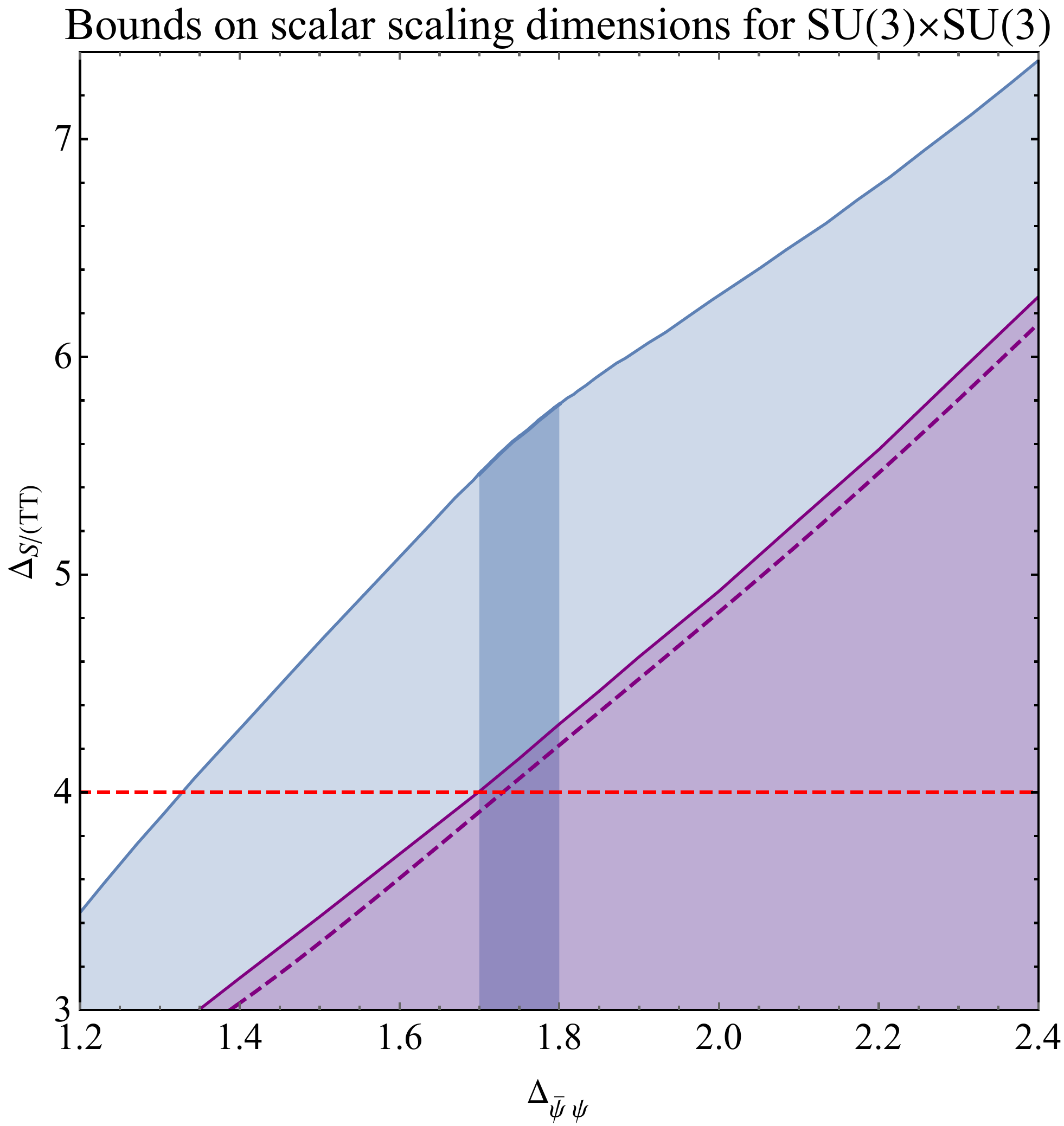}
 \begin{flushright}
\caption{Bounds on the scaling dimensions of the lowest scalars in the $SU(3)\times SU(3)$ singlet (blue line, $\Lambda=35$)  and symmetric - symmetric ($TT$) representations (purple line, $\Lambda=31$).
The singlet bound coincides with the bound on the $SO(18)$ singlet, while the $TT$ bound is slightly higher than the bound on the $SO(18)$ traceless symmetric scalar (dashed purple line, $\Lambda=31$). There is a mild kink in the singlet bound near $\Delta_{\bar{\psi}\psi}\in (1.7, 1.8)$ (dark blue shadowed region). The red dashed line gives the marginal condition $\Delta=4$.
  } \label{ON18}
\end{flushright}
\end{figure}

\begin{figure}[!htb]
\includegraphics[scale=0.65]{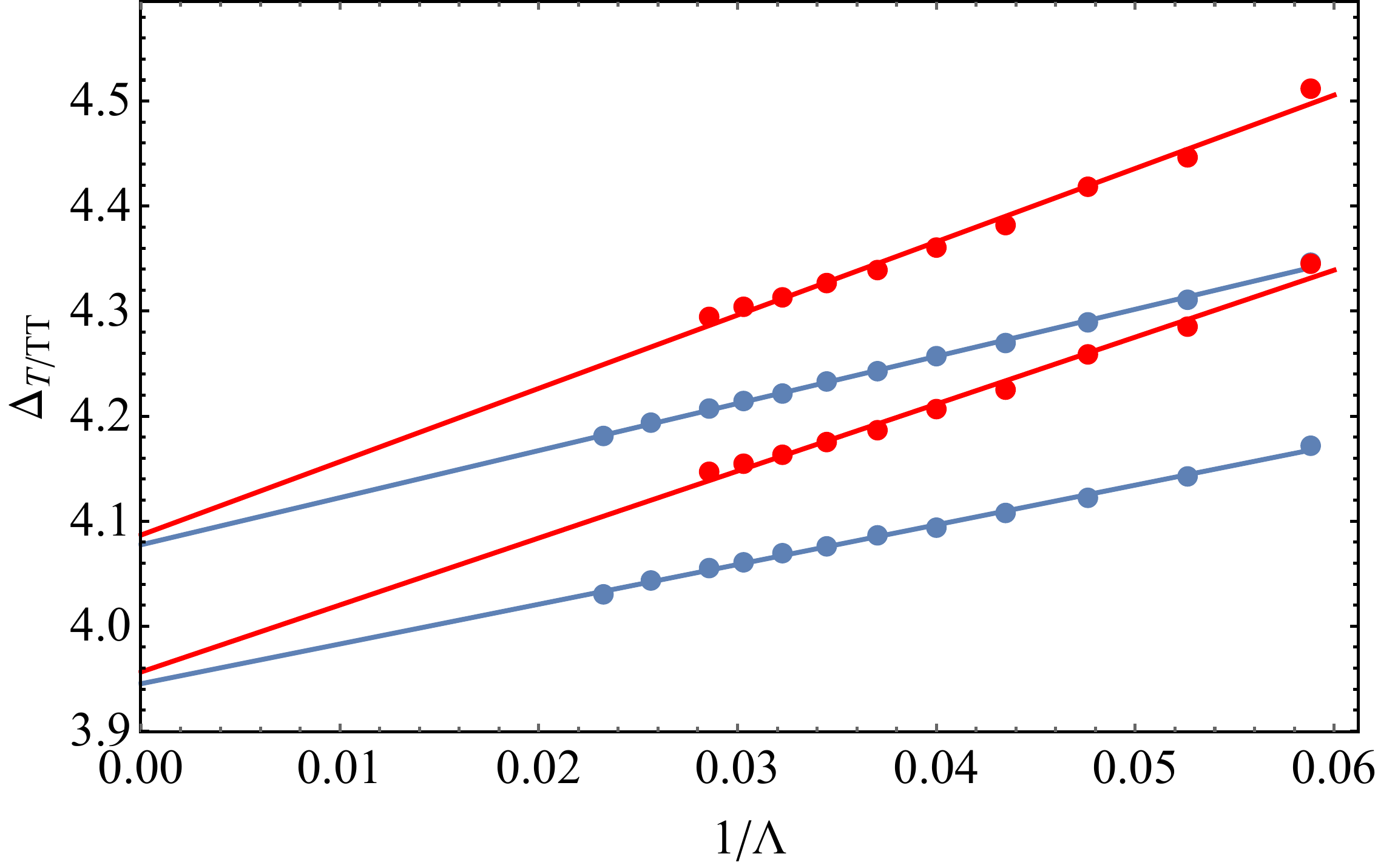}
 \begin{flushright}
\caption{The red (blue) lines are linear extrapolations of the upper bounds of lowest scalar in the symmetric-symmetric (traceless symmetric) sector of the $SU(3)\times SU(3)$ bi-fundamental ($SO(18)$ vector) bootstrap. The higher (lower) two lines correspond to $\Delta_\phi=1.80$ ($\Delta_\phi=1.75$). The points with $\Lambda=39, 43$ for the $SU(3)\times SU(3)$ bi-fundamental bootstrap are absent as they are fairly computationally intensive. 
  } \label{SU3extrp}
\end{flushright}
\end{figure}

In Figure \ref{ON18} we show bootstrap bounds on the scaling dimensions of the lowest scalars in $SU(3)\times SU(3)$ singlet and TT (symmetric-symmetric) representations, as well as bounds for $SO(18)$ singlets and symmetric tensors. In the singlet bounds (which coincide), there is a mild kink near the region $\Delta_{\bar{\psi}\psi}\sim (1.7, 1.8)$.\footnote{The kink becomes weaker near the critical flavor number $N^*$ and one can only identify a transition region in the bootstrap bound.
The operator decoupling in the singlet sector appears near $\Delta_\phi=1.75$ with $\Lambda=31$.
A better estimation on $\Delta_{\bar{\psi}\psi}$ could likely be obtained with higher numerical precision or the addition of gaps that create a sharp feature.} 
Based on the current numerical precision, we cannot make a reliable conclusion if the bound has two nearby kinks (as seems to occur in the analogous 3D bounds \cite{Li:2018lyb}). It would be interesting to check whether this is the case at higher numerical precision. 

 To study the convergence with $\Lambda$, we show linear extrapolations of the bootstrap bounds at $\Delta_{\phi} = (1.75, 1.8)$ in Figure \ref{SU3extrp}. 
The large $\Lambda$ limit of the $SU(3)\times SU(3)$  symmetric-symmetric scalar bound is quite close to the $SO(18)$ traceless symmetric scalar bound, though there is a notable distance between the two bounds at $\Lambda=31$. A trivial consequence is that the $SU(3)\times SU(3)$ symmetric-symmetric scalar bound also crosses marginality in the range $\Delta_\phi\in(1.75, 1.80)$.

As shown in Figure \ref{4DONs}, the bootstrap results suggest that the critical flavor number $N^*$ is below $N^*<N= 18$. 
Due to the ``fake" symmetry enhancement effect from the bootstrap algorithm, the value $N^*$ should be considered as describing the dimension of the representation of the external scalar for the symmetry ${\cal G}\subset SO(N^*)$.

The flavor number $N=18$ corresponds to a small value of $N_f$ when interpreted in terms of an $SU(N_f) \times SU(N_f)$ chiral symmetry. For example, the singlet bound from the $SO(18)$ vector bootstrap coincides with that from the $SU(3)\times SU(3)$ bi-fundamental bootstrap. Here the flavor number $N_f=3$ should be far below the lower bound of the conformal window of $SU(N_c)$ gauge theories with $N_c\geqslant3$ coupled to fundamental massless fermions. The chiral symmetry of $SU(2)$ gauge theories coupled to fundamental massless fermions is enhanced from $SU(N_f)\times SU(N_f)$ to $SU(2N_f)$ of which the fermion bilinears furnish a rank $2$ representation. According to our previous result, a ``fake" symmetry enhancement in the bootstrap bound of such fermion bilinears is expected as well, and the number of degrees of freedom of the fermion bilinears near/in the conformal window, such as $N_f=6,8$ \cite{Leino:2018yfd}, is significantly larger than $18$. The conclusion is that the kinks near $N^*$ are unlikely to be realized by $SU(N_c)$ gauge theories coupled to fundamental fermions. 

Instead, theories possessing a conformal window with small $N_f$ can potentially be obtained using fermions carrying multiple color indices \cite{Dietrich:2006cm}. In \cite{Ryttov:2010iz, Ryttov:2017kmx} the authors studied $SU(N_c)$ gauge theories coupled to massless fermions in the symmetric or anti-symmetric representations. The upper bounds on the conformal windows for rank $2$ representations are determined by the disappearance of asymptotic freedom
\beq
N_{f,\text{max}}=\frac{11N_c}{2(N_c\pm2)},
\eeq
with the $\pm$ respectively corresponding to symmetric and anti-symmetric representations. 
On the other hand, the lower bounds on the conformal windows are difficult to compute. Perturbatively, they can be estimated from a loop expansion of the beta function, where the leading terms are  \cite{Ryttov:2010iz}
\beq
N_{f,\text{min}}\sim \frac{17}{8}\mp 	\frac{323}{64}\frac{1}{ N_c}+\dots ~.
\eeq
The above perturbative results on $N_{f,\text{min}}$ need to be treated carefully. However, it is clear that gauge theories with rank $2$ representations can realize fixed points with small flavor numbers. In the strongly coupled region, the scaling dimensions of the fermion bilinears may obtain large corrections and have the possibility to be comparable with our bootstrap results. 
We postpone exploring explicit constructions of the possibilities for such theories to future work.

The kinks near $N^*$, if they relate to full-fledged CFTs, provide exceptional opportunities to study the critical behavior when the fixed point disappears, i.e., the mechanism by which conformality is lost, in a {\it nonperturbative} way.  As shown in Figure \ref{ON18}, for $N=18$ the scaling dimension of the fermion bilinears is $\Delta_{\bar{\psi}\psi}\sim (1.7, 1.8)$ at the kink, corresponding to a fermion anomalous mass dimension $\gamma_m=3-\Delta_{\bar{\psi}\psi}\sim (1.2, 1.3)$.
An open question relating to the loss of conformality in the CBZ fixed points is the anomalous mass dimension $\gamma_m$ near the lower bound of the conformal window. The unitarity bound requires $\gamma_m \leqslant 2$.
In the Veneziano limit,  $N_c\rightarrow \infty$ with fixed ratio $x=\frac{N_f}{N_c}$,
the anomalous mass dimension is expected to be $\gamma_m=1$. In this limit the loss of conformality  at the lower end of the conformal window $x=x^*$ is usually ascribed to an irrelevant four-fermion operator crossing marginality $\Delta_{(\bar{\psi}\psi)^2}=4$. In the Veneziano limit we have the factorization $\Delta_{(\bar{\psi}\psi)^2}=2\Delta_{\bar{\psi}\psi}$, which gives $\gamma_m=1$ at $x^*$.
A similar bound on the anomalous mass dimension can also be obtained from the Schwinger-Dyson equation \cite{Appelquist:1988yc}. The anomalous dimension $\gamma_m$ necessarily receives corrections at finite $N_f$, and our results near $N^*$ may provide evidence of a large fermion anomalous mass dimension $\gamma_m > 1$.

According to the results shown in Figures \ref{4DONs} and \ref{ON18}, near the critical flavor number, depending on the exact value of $N^*$, there could be two quite different scenarios for the disappearance of the kinks. 
For $N^*\sim 18$, near the kink with $\Delta_{\bar{\psi}\psi}\sim (1.7, 1.8)$, the upper bound on the scaling dimension of the  symmetric-symmetric scalar gets close to marginal. In other words, it suggests that the disappearance of the kinks may coincide with an irrelevant scalar  in the $V_{T,T}^{(+)}$ sector crossing marginality.\footnote{Bounds on the scaling dimensions of scalars in other sectors are higher. However, in certain sectors they are close to the bound of $V_{T,T}^{(+)}$ and they may also play an important role near $N^*$. We leave a more comprehensive analysis of these sectors for a future study.}  In contrast, for $N^*\sim14$, it is the singlet scalar that crosses marginality. Both of the scenarios suggest that the disappearance of the kinks is accompanied by a marginal scalar in some representation.

The appearance of a marginal scalar near $N^*$ is nicely consistent with the scenario of merger and annihilation of fixed points \cite{10.1143/PTP.105.809,PhysRevB.71.184519,Gies:2005as,Kaplan:2009kr, Gorbenko:2018ncu}, which was proposed as a scenario explaining how conformality is lost in the CBZ fixed points near $N^*$.
In \cite{Kaplan:2009kr} the authors suggested that near the lower bound of the conformal window, the CBZ fixed points approach another UV fixed point, QCD$^*$, generated by a certain irrelevant operator (e.g.~a scalar built out of four fermions). At the critical flavor number $N^*$, the irrelevant operator becomes marginal, which further drives the two fixed points to merge with each other and triggers chiral symmetry breaking. The two fixed points disappear for $N<N^*$. 
A more  comprehensive study on how conformality could be lost is provided in \cite{Gorbenko:2018ncu}, which analyzed additional scenarios for the behavior of fixed points near $N^*$. In general, the loss of conformality could possibly be triggered by a non-singlet scalar reaching marginality and the two lines of IR and UV fixed points may cross instead of merge with each other, depending on the symmetry of the marginal operator. The evolution of fixed points can be quantitatively described in the context of bifurcation theory \cite{Gukov:2016tnp}. 

A solid prediction of this mechanism is that an irrelevant operator should become marginal at $N^*$. 
 In particular, the authors of  \cite{Kaplan:2009kr} expected the loss of conformality is triggered by a marginal singlet scalar. This could be consistent with the bootstrap results given $N^*\simeq 14$, in which the kink disappears when the lowest singlet approaches marginality. However, in another scenario with $N^*\simeq 18$, the lowest singlet remains irrelevant while the symmetric-symmetric scalar (or traceless symmetric scalar in an $SO(N)$ symmetric theory) approaches marginality. If the loss of conformality is triggered by a non-singlet scalar, then it would indicate a scenario of conformality being lost different from that proposed in  \cite{Kaplan:2009kr}. Here a precise estimation on the critical flavor number $N^*$ is of crucial importance to determine which scenario is favored by the bootstrap results. We expect that the connection between operators decoupling and the kink in the bootstrap bound could be helpful for solving this problem, given it can be established in a more numerically reliable fashion.

According to the merger and annihilation mechanism, regardless of the details of the deformation, there should be a nearby UV fixed point generated by the approximately marginal operator.  We do not see clear evidence of the existence of a nearby kink in the bootstrap bound. This is different from the 3D bootstrap results, in which the singlet bound shows two nearby kinks above the critical flavor number $N^*$  \cite{Li:2018lyb}.  It would be interesting to know if this is just due to the issue of numerical convergence of the bootstrap bounds in 4D, or there is a more fundamental reason for its absence.

\FloatBarrier\subsection{Bounds with $N_f=12$}
The kinks appearing in the $SU(N_f)\times SU(N_f)$ singlet bounds correspond to a series of non-trivial solutions to the crossing equations, which may relate to full-fledged CFTs. The only known candidates for 4D non-supersymmetric CFTs are the fixed points of gauge theories, like CBZ fixed points. It is of great interest to compare bootstrap results with the CFT data of CBZ fixed points obtained from other approaches. Unfortunately, our knowledge on CBZ fixed points in the strong-coupling limit is quite limited. One of the best studied theories of this type is the $SU(3)$ gauge theory with $12$ flavors of massless fermions in the fundamental representation.
There is strong evidence from lattice simulations that this theory admits an IR fixed point \cite{Appelquist:2007hu}.
The fermion  anomalous mass dimension $\gamma_m=3-\Delta_{\bar{\psi}\psi}$ is particularly easy to access in the lattice studies.  
The estimated values of $\gamma_m$ spread in a range: $\gamma_m\sim 0.403(13)$ \cite{Appelquist:2011dp},  $\gamma_m\sim 0.35$ \cite{DeGrand:2011cu},  $\gamma_m\in [0.3, 0.4]$ \cite{Aoki:2012eq}, $\gamma_m\sim 0.32(3)$ \cite{Cheng:2013eu},  $\gamma_m\sim 0.25$ \cite{Hasenfratz:2013eka}, $\gamma_m\sim 0.235(15)$ \cite{Hasenfratz:2016dou}, $\gamma_m\sim 0.235(46)$ \cite{Lombardo:2014pda}, $\gamma_m\in [0.2, 0.4]$  \cite{Fodor:2012et},  and $\gamma_m\sim 0.23(6) $ \cite{Carosso:2018bmz}.
The anomalous mass dimension has also been computed using perturbative approaches: $\gamma_m\sim 0.25$ at four-loops \cite{Chetyrkin:1997dh, Vermaseren:1997fq}, $\gamma_m\sim 0.255$ at five-loops \cite{Ryttov:2016ner}, $\gamma_m\sim 0.338$ at fourth-order using scheme-independent series expansions \cite{Ryttov:2016asb,Ryttov:2017kmx}, $\gamma_m\sim 0.3375 - 0.352$ using Pad\'e resummations of these expansions \cite{Ryttov:2017lkz}, and $\gamma \sim 0.320(85)$ in a recent work based on a Pad\'e-Borel resummation of the coupling expansion \cite{DiPietro:2020jne}.

\begin{figure}[!htb]
\includegraphics[scale=0.6]{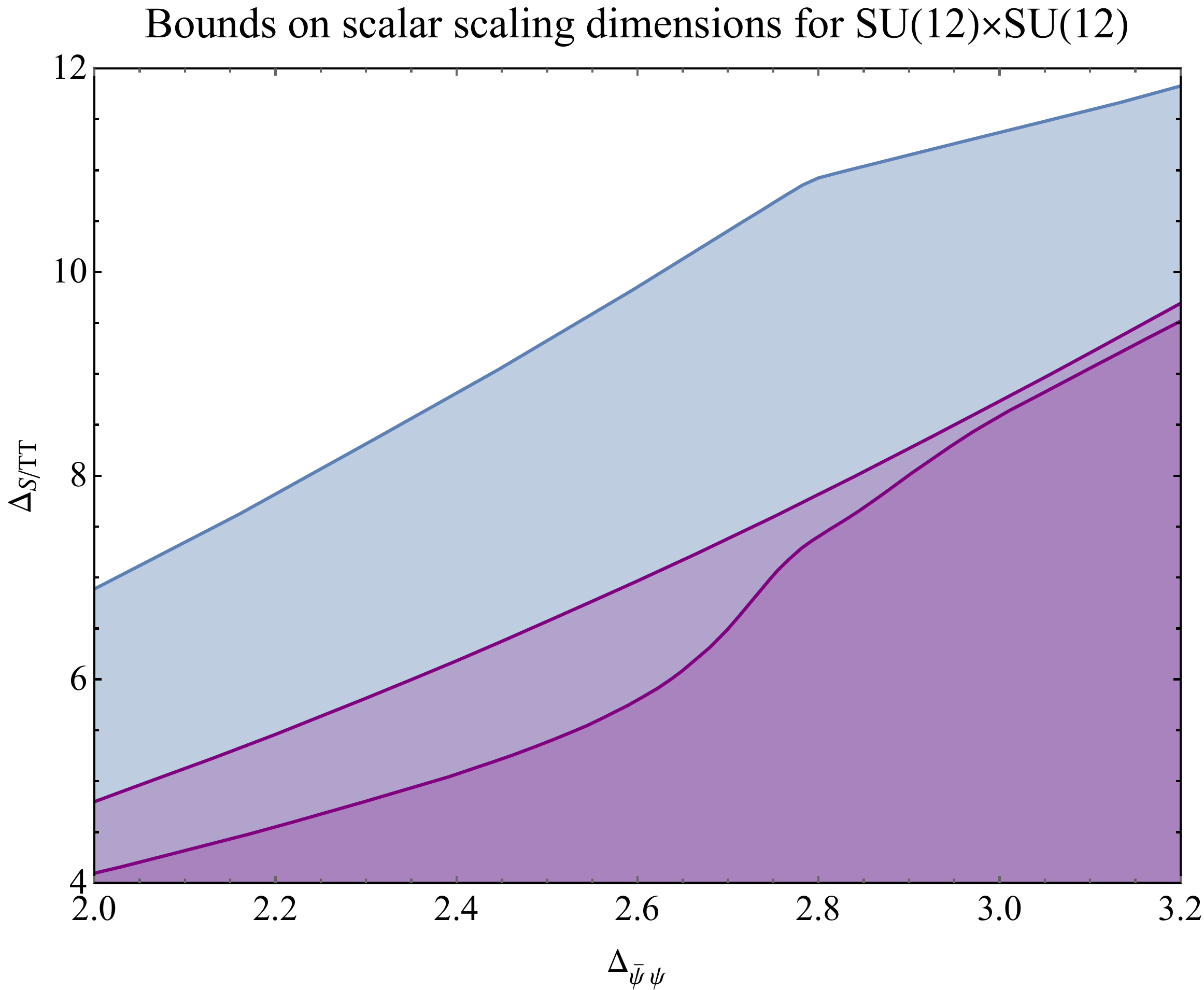}
 \begin{flushright}
\caption{Bounds on scaling dimensions of $SU(12)\times SU(12)$ singlet (blue line, $\Lambda=43$) and  symmetric-symmetric ($TT$) (higher purple line, $\Lambda=31$) scalars. The bound on the scaling dimension of the lowest traceless-symmetric scalar from the $SO(288)$ vector bootstrap is also presented (lower purple line, $\Lambda=31$), which is identical to the  bound on the $TT$ scalar from the $SU(12)\times SU(12)$ bi-fundamental bootstrap with an extra assumption that the lowest non-singlet scalars appearing in the crossing equation (\ref{4DSUN2ce}) have the same scaling dimension. This assumption is true in the large $N_f$ limit and violated by $1/N_f$ corrections.
  } \label{SU(12)}
\end{flushright}
\end{figure} 

The bootstrap results from the crossing equation (\ref{4DSUN2ce}) with flavor number $N_f=12$ are provided in Figure \ref{SU(12)}. Bounds on the scaling dimensions of the lowest scalars in the $V_{S,S}^{(+)}$ and $V_{T,T}^{(+)}$ sectors of the crossing equation (\ref{4DSUN2ce}) are shown in the figure. We also provide the bound on the scaling dimension of the lowest scalar in the $V_{T}^{(+)}$ sector of the crossing equation (\ref{ONce}), which we have argued can give a leading-order approximation to the bound on the the four-fermion operators, up to $1/N_f$ corrections in different sectors. The upper bound changes notably by increasing $\Lambda$, suggesting that current bound is not close to the optimal bound and can not be used to estimate the scaling dimension of the lowest singlet scalar. 

 We tried to use extrapolation to study the large $\Lambda$ behavior of the bound. However, the $\Delta \sim 1/\Lambda$ relation is not linear: the singlet upper bound near the kink decreases from 16.6 ($\Lambda=19$) to 11.9 ($\Lambda=27$), 11.4 ($\Lambda=31$), 11.1 ($\Lambda=35$), 10.9 ($\Lambda=39$) and 10.8 ($\Lambda=43$). The bounds with larger $\Lambda\geqslant31$ are more close to a linear relation $\Delta_S \propto 1/\Lambda$ and the extrapolation based these points gives $\Delta_S\sim 9.2$. Nevertheless, there are considerable uncertainties in estimating the positions of the kinks and a significant correction to this extrapolation is expected.  Comparing with the extrapolations of the bootstrap bounds with small $N$ (Fig.~\ref{SOextrp}), the convergence of the singlet bootstrap bounds with larger $N$ appears more subtle.  Note that in the large $N$ limit, the OPE coefficients of (non-unit) singlet operators are usually $1/N$ suppressed, while for the operators in the non-singlet representations, such as the symmetric-symmetric ($TT$) representation, their OPE coefficients are of $O(1)$. The small OPE coefficients of singlet operators could make it harder for the numerical bootstrap to capture the singlet spectrum. Besides the convergence issue, it would also be interesting to know if the singlet upper bound could be modified by other effects, like the mixing of spectra seen in the $SO(N)$ vector bootstrap bound on the $T$ scalar at large $N$  (Fig.~\ref{EFMinf}).
 
Nevertheless, the $x$-position of the kink, which corresponds to the scaling dimension of the  $SU(12)\times SU(12)$ bi-fundamental, does not change significantly by increasing $\Lambda$. Assuming this continues to hold at even larger values of $\Lambda$, it suggests that the $x$-position of the kink in the optimal bound could be near the current value $\Delta_{\bar{\psi}\psi}\sim 2.78$, i.e., the fermion anomalous mass dimension is near $\gamma_m\sim 0.22$! This is comparable to the estimates of the anomalous mass dimension of $12$ flavor QCD using other approaches. 
However, in $12$ flavor QCD, the lowest parity even singlet scalar that can appear in the fermion bilinear OPE  $\cO\times\cO^\dagger$   is expected to be the gauge singlet $\mathrm{Tr}[F_{\mu\nu}F^{\mu\nu}]$. Its anomalous dimension $\gamma_g\equiv \Delta_{F^2}-4$ has been computed using lattice simulations ($\gamma_g=0.26(2)$) \cite{Hasenfratz:2016dou} and also the perturbative approach ($\gamma_g=0.23(6)$)  \cite{DiPietro:2020jne}, which is way below our current upper bound. A better understanding and numerical control on the singlet bounds would be especially important to study the possible	 connections between these kinks and the CBZ fixed points.

It will be interesting in future work to seek reliable estimates of scaling dimensions of other sectors of this theory and compare them to known gauge theories. As already mentioned, the present bootstrap implementation does not include any direct information on the gauge interactions, such as the underlying gauge group or how the fermions transform under the gauge symmetry. $SU(N_c)$ QCD with $12$ fundamental massless fermions is one of the candidates, while there could be many other candidates with different gauge symmetries (e.g.~$SO(N_c)$ or $Sp(N_c)$ \cite{Sannino:2009aw,Mojaza:2012zd,Ryttov:2017dhd}, or semi-simple gauge groups~\cite{Bond:2017lnq}) coupled to matter in various types of representations. This problem might be partially resolved by bootstrapping mixed correlators containing the $SU(3)$ baryon operator,
\beq
\cO^{ijk}=\epsilon_{\alpha\beta\gamma}\psi^{i\alpha}\psi^{j\beta}\psi^{k\gamma},
\eeq
where $\{\alpha,\beta,\gamma\}$ ($\{i,j, k\}$) are the color (flavor) indices.
The operator $\cO^{ijk}$ is not gauge invariant in $SU(N_c)$ gauge theories with $N_c\neq 3$ and could help to distinguish this particular theory from many other candidates. This operator is fermionic and lives in a nontrivial representation of the flavor symmetry. The 4D fermion bootstrap has been studied in \cite{Karateev:2019pvw} and the techniques developed in the work can be directly employed for this study.\footnote{The 3D fermion bootstrap has been explored in \cite{Iliesiu:2015qra, Iliesiu:2017nrv} which show kinks corresponding to the Gross-Neveu-Yukawa models.} The baryon operator has UV scaling dimension $\Delta_{B}^{UV}=9/2$, while it receives significant corrections due to strong coupling, and its IR scaling dimension $\Delta_{B}^{IR}$ could possibly be much lower, see \cite{Gracey:2018oym} for a scheme-independent perturbative study on the anomalous dimensions of baryon operators. The magnitude of $\Delta_{B}^{IR}$ will likely directly affect whether the numerical bootstrap can generate strong bounds towards a physical theory. The main technical obstacle is due to the global symmetry and spinor indices, which lead to quite cumbersome crossing equations. Recent developments in numerical bootstrap techniques \cite{Go:2019lke, Landry:2019qug, Chester:2019ifh, Erramilli:2019njx} will be helpful for pursuing this study.

\FloatBarrier\section{Kinks in the 5D $SO(N)$ vector bootstrap}
\label{sec:5d}

\begin{figure}[!htb]
\includegraphics[scale=0.7]{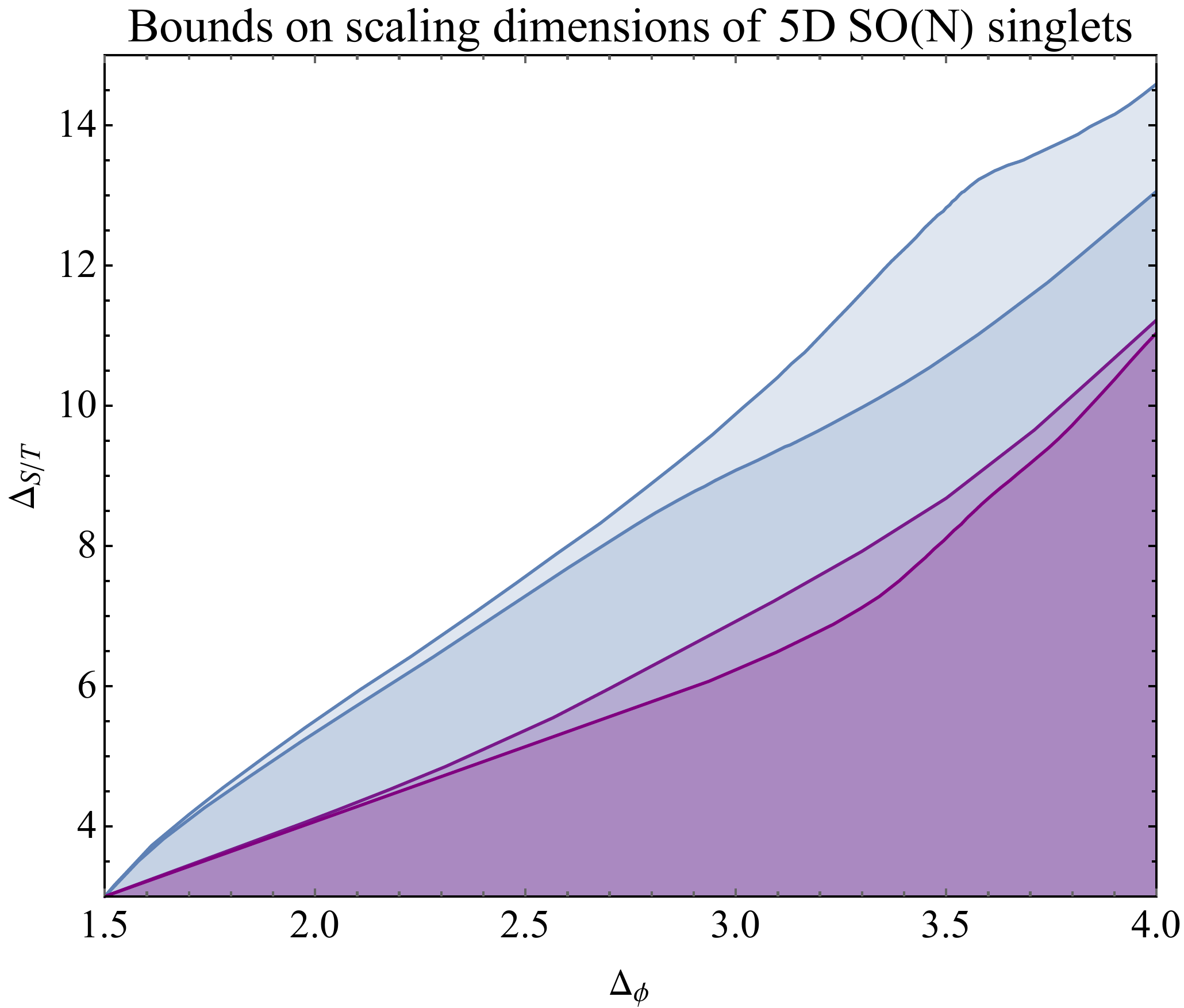}
 \begin{flushright}
\caption{Bounds on the scaling dimensions of 5D $SO(60/500)$ singlet (blue lines) and  traceless symmetric (purple lines) scalars. The higher blue line and lower purple line correspond to $N=500$.  The bounds are estimated with $\Lambda=31$. 
  } \label{5DON}
\end{flushright}
\end{figure}

Finally we end with some comments on the situation in 5D. Recall that the kinks ${\cal T}_D$ which approach free fermion theory in the large $N$ limit appear for general dimensions D. In analogy with the bootstrap results in 3 and 4 dimensions, in 5D one might expect a similar infinite family of kinks with a modest higher value of critical number $N^*$. On the other hand, it is known that  non-supersymmetric unitary interacting CFTs are rare in 5D, see \cite{BenettiGenolini:2019zth} for a recent study. The standard approach to constructing an interacting field theory is to deform a UV Gaussian fixed point with relevant operators. However, in higher dimensions $D\geqslant5$, there are only a few relevant operators in a free fermion theory and it is difficult to realize IR fixed points using a Lagrangian approach.
From this point of view, bootstrap results in  5D are likely to be qualitatively different from the results in lower dimensions.
Actually we find there is no sharp kink in the singlet bounds of the $SO(N)$ vector bootstrap for $N\leqslant60$. Notable kinks in the singlet bound do appear for large $N$, e.g.,  $N=500$, though we do not know if the kinks remain in the bounds with higher numerical precision. The kinks in 5D seem to have a notably larger critical number $N^*$ than their lower dimension analogs.

It is interesting to compare our results with previous studies on the 5D ``cubic" model \cite{Fei:2014yja, Fei:2014xta} with bosonic interactions $\mathcal{L} \sim \sigma \phi^{i} \phi^{i}$, which, in the large $N$ limit, approaches another theory with conserved higher spin currents: the ($N=\infty$) critical boson theory.
The cubic model provides a UV-complete version of the 5D dimensional continuation of the critical $O(N)$ vector models in $2< D<4$. Perturbatively the theory is  stable and unitary  for sufficiently large $N$ above a critical value  $N>N^*$. Remarkably, the bootstrap approach can provide sharp results on the critical 5D cubic models which are consistent with the large $N$ or $(6-\epsilon)$-expansion perturbative predictions \cite{Nakayama:2014yia, Bae:2014hia, Chester:2014gqa, Li:2016wdp, Gracey:2015tta}.\footnote{Note that the critical boson has scaling dimension $\Delta=2+O(1/N)$, which is actually lower than that of the 5D free boson bilinear operators  $\Delta=3$. Therefore we cannot the kinks corresponding to these theories in Figure \ref{5DON} unless we impose gaps in certain sectors which carve out regions below the free boson theory.}
On the other hand, it turns out to be quite subtle to determine the critical number $N^*$. Especially in the mixed correlator bootstrap \cite{Li:2016wdp}, for relatively small $N^*\simeq100$ the physically allowed CFT data can be isolated into a small island with suitable assumptions, while after increasing numerical precision the island disappears. An explanation of this phenomena is that there are small non-unitary effects in the CFT data which can only be seen using the conformal bootstrap with sufficiently high precision. Remarkably, such non-unitary effects have been found to be generated by instantons and can be computed analytically \cite{Giombi:2019upv}.

One of the motivations of the cubic model \cite{Fei:2014yja, Fei:2014xta} is to construct 5D theories with slightly broken higher spin symmetry. The cubic model, in the $(6-\epsilon)$-expansion description,  corresponds to a free boson theory perturbed by cubic interactions.
Theories with exact higher spin symmetry can also be realized with free fermions, and one may alternatively construct theories with slightly broken higher spin symmetry by perturbing free fermion theory. In 3D, these two different types of theories have been partially revealed in bootstrap results: we have two families of prominent kinks which respectively approach the critical boson or free fermions in the large $N$ limit \cite{Li:2018lyb}.

Inspired by the 5D cubic model and bootstrap results, we may ask this question: can we construct 5D theories as deformations of free fermion theory which admit IR fixed points?
The theory could be unitary and stable perturbatively for sufficiently large flavor number. One might expect that the deformations correspond to adding gauge interactions, as is suggested by the lower-dimension bootstrap results. This is also evident for a simple reason that in a non-gauged fermionic theory, local interactions with fermions are strongly UV irrelevant in higher dimensions $D\geqslant 5$.
A natural approach to realize 5D UV fixed points by perturbing free fermion theory  is to take the $D=4+\epsilon$ dimensional continuation of 4D gauge theories, e.g., QED coupled to fermions. 

In the $(4+\epsilon)$-expansion, at the one-loop level QED in $D>4$ admits a UV fixed point. However, the gauge coupling is imaginary at the fixed point so the theory is non-unitary in a certain neighborhood of $D=4$. In \cite{Giombi:2015haa}
the authors proposed a UV completion in terms of a higher-derivative renormalizable Abelian gauge theory in $D=6-\epsilon$ dimensions, in which the CFT is realized as an IR fixed point. Unfortunately, the IR fixed point remains non-unitary at the one-loop level. It is not clear if the fixed points could be perturbatively stable and unitary with $\epsilon=1$. Perhaps one can construct more interesting fixed points using non-Abelian gauge theories, which have a chance to be perturbatively stable and unitary.\footnote{Fixed points of 6D QCD have been studied in \cite{Gracey:2015xmw} using a perturbative approach, which suggests a conformal window analogous to 4D QCD. We also observed kinks in the bounds from the 6D $SO(N)$ vector bootstrap, but it requires a careful study before any solid conclusion can be reached.}

\FloatBarrier\section{Discussion}
\label{sec:discussion}

In this work we have explored the properties of certain prominent discontinuities in the conformal bootstrap bounds with global symmetries in general dimensions. In the 3D $SO(N)$ vector bootstrap, there are two infinite families of kinks in the bounds on scaling dimensions of $SO(N)$ singlets, which, in the large $N$ limit, respectively approach solutions with an $SO(N)$ vector $\phi_i$ of scaling dimension $\Delta_{\phi}=D/2-1$ and $\Delta_{\phi}=D-1$, along with a series of conserved higher spin currents.  The two families of kinks at finite $N$ are expected to be interacting perturbations of free boson and free fermion theories. 
The kinks near the unitary bound $\Delta_\phi\sim D/2-1$ relate to the critical $O(N)$ vector models \cite{Kos:2013tga}, while another family of two adjacent kinks ${\cal T}_{3D}$ with large anomalous dimensions are potentially related to the IR fixed points of QED$_3$ and QED$_3$-GNY.
In 4D there are no kinks related to deformations of free boson theory. This is expected since there is no interacting critical $O(N)$ vector model in 4D. The kinks  ${\cal T}_{4D}$ deformed from free fermions remain in 4D and we speculate that they correspond to fixed points of gauge theories coupled to multi-flavor fermions.
In 5D the kinks deformed from free bosons reappear in the numerical bootstrap bounds at large $N$, and relate to the IR fixed points of 5D cubic models. The kinks ${\cal T}_{5D}$ also persist with sufficiently large flavor number.

We also studied the coincidences between bootstrap bounds on scaling dimensions assuming different global symmetries.  Generically, for the single correlator bootstrap with an external scalar in a representation $\cal R$ of a group $\cal G$ with dimensionality $\cal N$, the bound on the scaling dimension of  the lowest singlet scalar appearing in the $\cal R\times \bar{\cal R}$ OPE is expected to coincide with the singlet bound from the $SO(\cal N)$ vector bootstrap. Moreover, we found a coincidence to also appear in the whole spectra of extremal solutions following the branching rules of $SO(\cal N)\rightarrow \cal G$. We provided an explicit proof for the bound coincidence between the $SO(2N)$ vector and $SU(N)$ fundamental bootstrap, by finding a transformation between the linear functionals for the $SO(2N)$ vector and $SU(N)$ fundamental crossing equations, which satisfy the required positivity conditions. The proof can be straightforwardly generalized to different global symmetries and it reveals an interesting connection between the structure of group representations and the bootstrap constraints.

We particularly focused on the 4D bootstrap results and their possible connections with the well known 4D CBZ fixed points of Yang-Mills theories coupled to massless fermions. These theories could have interesting applications to physics beyond the Standard Model or nonperturbative approaches to ordinary QCD (such as Hamiltonian truncation). Since the new family of kinks seem to correspond to deformations of free fermion theory, also inspired by their 3D analogs, we speculate these kinks correspond to fermionic theories with gauge interactions, of which the CBZ fixed points are known candidates. We discussed possible field realizations of the kinks near the apparent critical flavor number, which corresponds to a small flavor number and a large anomalous mass dimension.

We showed that the bootstrap results may help us to understand how conformality is lost near the lower bound of the conformal window. However, a solid conclusion on this seems to require a more precise estimation of the critical flavor number $N^*$.
We also presented in detail the bootstrap results with flavor number $N_f=12$, which show additional promising comparisons with lattice and perturbative results. It is expected that more information on the precise nature of the underlying theories can be revealed by bootstrapping mixed correlators with meson operators, conserved currents, and possibly baryon operators (e.g., in $SU(3)$ gauge theories). 

The 5D $SO(N)$ vector bootstrap results are reminiscent to those in 3D, where there are two family of kinks respectively approaching free boson and free fermion bilinears in the large $N$ limit. 
The kinks deformed from free boson theory are the dimensional continuation of the critical $O(N)$ vector model in lower dimensions, for which the cubic model provides a UV complete description. 
However, it has been shown that the IR fixed points of the 5D cubic model, though perturbatively stable and unitary with sufficient large $N$, contains non-unitary factors due to instanton corrections which are too small to be detected by previous numerical precision. Our bootstrap results suggest there is also a family of CFTs deformed from free fermion theory at large $N$, which seems to be unitary at the current numerical precision. Inspired by the cubic model, it would be very interesting to know if they admit a UV complete Lagrangian description in which unitarity is violated by nonperturbative effects only. 

The new kinks that appear to correspond to deformations of free fermion theories are welcome surprises for the conformal bootstrap. This is especially encouraging for the possibility to bootstrap fermionic gauge theories with strong interactions, which play critical roles in many applications but require nonperturbative treatments. One obstacle towards a precise estimation of the CFT data of the underlying theories is that the singlet bound converges rather slowly and is not yet close to its optimal solution. In the limit $N\rightarrow \infty$, the singlet upper bound also tends to disappear. At finite but large $N$, the OPE coefficients in the singlet sector are $1/N$ suppressed so it becomes challenging for the numerical bootstrap to capture these small factors.\footnote{On the other hand, it is surprising that sharp kinks can still be formed with stable $SO(N)$ vector or fermion bilinear scaling dimensions.} The problem remains but is less severe at the more interesting (for applications) case of small $N$ or flavor number, where the theories are expected to be more strongly coupled. Bounds on the scaling dimensions of certain non-singlet scalars perform better in this aspect. In comparison with the 4D CBZ fixed points, conformal QED$_3$ provides a relatively simple laboratory for understanding these issues. We expect that a precision bootstrap study of mixed correlators in QED$_3$ can help to illustrate an effective approach for evaluating the CFT data of fermionic gauged CFTs.  We hope to report developments on this problem in the near future \cite{AELP}.

\section*{Acknowledgements}

We thank Soner Albayrak, Thomas Appelquist, Shai Chester, Rajeev Erramilli, George Fleming, Holger Gies, John Gracey, Yin-Chen He, Luca Iliesiu, Denis Karateev, Petr Kravchuk, Walter Landry, Daniel Litim, Junyu Liu, Junchen Rong, Slava Rychkov, Robert Shrock, Marco Serone, David Simmons-Duffin, Ning Su, and Alessandro Vichi for discussions.  We are grateful to Slava Rychkov for valuable comments on the draft. ZL would like to thank Miguel Costa and Christopher Pope for their support during early stage of this work. ZL also thanks Ning Su for help with Simpleboot. The work of ZL and DP is supported by Simons Foundation grant 488651 (Simons Collaboration on the Nonperturbative Bootstrap) and DOE grant no.\ DE-SC0020318. Computations in this work were carried out on the Yale Grace computing cluster, supported by the facilities and staff of the Yale University Faculty of Sciences High Performance Computing Center, and the Mac Lab cluster supported by the Department of Physics and Astronomy, Texas A\&M University.


\appendix
\section{Crossing equations for the $SU(N_f)\times SU(N_f)$ bi-fundamental bootstrap}
\label{app:crossing}
4D QCD coupled to $N_f$ flavor massless fermions has chiral symmetry $SU(N_f)\times SU(N_f)\times U(1)_V$. The chiral symmetry is unbroken in the IR conformal phase, i.e., at the CBZ fixed points.
The gauge-invariant fermion bilinears $\cO_{\text{bf}}=\bar{\psi}\psi$ construct bi-fundamental representations of this chiral symmetry. The 4-point correlator $\langle\cO_{\text{bf}}(x_1) \cO^\dagger_{\text{bf}}(x_2)\cO_{\text{bf}}(x_3) \cO^\dagger_{\text{bf}}(x_4)\rangle$ provides a natural candidate for a bootstrap study of this theory. The explicit form of the crossing equation for this correlator has been computed in \cite{Nakayama:2016knq}. It can be written in a compact form, as shown in (\ref{4DSUN2ce}) and below,
\bea
\sum_{O\in\cO\times\cO^\dagger}\lambda^2_O V_{S,S,\Delta,\ell}^{(\pm)}+\sum_{O\in\cO\times\cO^\dagger}\lambda^2_O V_{\text{Adj,Adj},\Delta,\ell}^{(\pm)}+\sum_{O\in\cO\times\cO^\dagger}\lambda^2_O V_{\text{Adj},S,\Delta,\ell}^{(\pm)}+\nn\\
\sum_{O\in\cO\times \cO}\lambda^2_O V_{T,T,\Delta,\ell}^{(+)}+\sum_{O\in\cO\times \cO}\lambda^2_O V_{T,A,\Delta,\ell}^{(-)}+ \sum_{O\in\cO\times \cO}\lambda^2_O V_{A,A,\Delta,\ell}^{(+)}+\cdots=0,
\eea
where the $V_X^{(\pm/+/-)}$ represent the following $9$-component vectors:

\begin{align}
V_{S,S,\Delta,\ell}^{(\pm)} &= \left( \begin{array}{cc}  F \\ H \\ 0 \\ 0 \\ 0 \\ (-1)^\ell F \\ (-1)^\ell H \\  0 \\ 0 \\
\end{array} \right) \ , \ \
V_{\text{Adj,Adj},\Delta,\ell}^{(\pm)} = \left( \begin{array}{cc}  (1+\frac{1}{N^2}) F \\ (-1+\frac{1}{N^2}) H \\ -\frac{2}{N}F \\ (-1)^\ell F \\ -(-1)^\ell H \\  (-1)^\ell\frac{1}{N^2} F \\ (-1)^\ell\frac{1}{N^2} H \\ -(-1)^\ell\frac{1}{N} F \\ -(-1)^\ell\frac{1}{N} H \\
\end{array} \right) \ , \ \
 V_{T,T,\Delta,\ell}^{(+)} =  \left( \begin{array}{cc}  0 \\ 0 \\ 0 \\ F \\ H \\ F \\ -H \\  F \\ -H \\
\end{array} \right)\ ,
\end{align}
\begin{align}
V_{\text{Adj},S,\Delta,\ell}^{(\pm)} &= \left( \begin{array}{cc}  -\frac{2}{N} F \\ -\frac{2}{N} H \\ 2 F \\ 0 \\ 0 \\  -(-1)^\ell\frac{2}{N} F \\ -(-1)^\ell\frac{2}{N} H \\  (-1)^\ell F \\ (-1)^\ell H \\
\end{array} \right)
 \ , \ \
V_{T,A,\Delta,\ell}^{(-)} =  \left( \begin{array}{cc}  0 \\ 0 \\ 0 \\ -F \\ -H \\ F \\ -H \\  0 \\ 0 \\
\end{array} \right) \ , \ \
V_{A,A,\Delta,\ell}^{(+)} = \left( \begin{array}{cc}  0 \\ 0 \\ 0 \\ F \\ H \\ F \\ -H \\  -F \\ H \\
\end{array} \right) \ . \ \
\end{align}
Here the conformal blocks $F/H$ are respectively defined through (\ref{Fblock}) and (\ref{Hblock}), and the $\cdots$ represent symmetrized/conjugate representations whose vectors take the same form as above.

\bibliography{refs}
\bibliographystyle{utphys}
\end{document}